\documentclass[final,5p,times,twocolumn]{elsarticle}

\usepackage{CJK}%
\usepackage{color}%
\usepackage{amsmath,bm}%
\usepackage{graphicx}%
\usepackage{dcolumn}% Align table columns on decimal point
\usepackage{longtable}
\usepackage{array}
\usepackage[switch,mathlines]{lineno}
\usepackage{ulem}
\usepackage{setspace}
\usepackage{subfigure}

\journal{Physical Review C}

\begin{document}
%\graphicspath{{figures/}}
\begin{frontmatter}

\title{Determining impact parameters of heavy-ion collisions at low-intermediate incident energies using deep learning with convolutional neural network}
\author[label1]{X. Zhang}
\author[label1]{Y. Huang}
\author[label1]{W. Lin}
\author[label1]{X. Liu\corref{cor1}}
\cortext[cor1]{Email address: liuxingquan@scu.edu.cn}
\author[label2]{H. Zheng}
\author[label3]{R. Wada}
\author[label3,label7]{A. Bonasera}
\author[label8]{Z. Chen}
\author[label1]{L. Chen}
\author[label1]{J. Han}
\author[label8]{R. Han}
\author[label5]{M. Huang}
\author[label8]{Q. Hu}
\author[label1]{Q. Leng}
\author[label4]{C. W. Ma}
\author[label1]{G. Qu}
\author[label1]{P. Ren}
\author[label8]{G. Tian}
\author[label1]{Z. Xu}
\author[label9]{Z. Yang}
\author[label10]{L. Zhang}

\address[label1]{Key Laboratory of Radiation Physics and Technology of the Ministry of Education, Sichuan University, Chengdu 610064,	China}
\address[label2]{School of Physics and Information Technology, Shaanxi Normal University, Xi'an 710119, China}
\address[label3]{Cyclotron Institute, Texas A$\&$M University, College Station, Texas 77843}
\address[label7]{Laboratori Nazionali del Sud, INFN,via Santa Sofia, 62, 95123 Catania, Italy}
\address[label8]{Institute of Modern Physics, Chinese Academy of Sciences, Lanzhou 730000, China}
\address[label5]{College of Physics and Electronics information, Inner Mongolia University for Nationalities, Tongliao, 028000, China}
\address[label4]{School of Physics, Henan Normal University, Xinxiang 453007, China}
\address[label9]{Science and Technology on Reactor System Design Technology Laboratory, Nuclear Power Institute of China, Chengdu 610213,  China}
\address[label10]{School of Information Sciences and Engineering, Lanzhou University, Lanzhou 730000, China}

\begin{abstract}
A deep learning based method with the convolutional neural network (CNN) algorithm for determining the impact parameters is developed using the constrained molecular dynamics model simulations, focusing on the heavy-ion collisions at the low-intermediate incident energies from several ten to one hundred MeV/nucleon in which the emissions of heavy fragments with the charge numbers larger than 3 become crucial. To make the CNN applicable in the task of the impact parameter determination at the present energy range, specific improvements are made in the input selection, the CNN construction and the CNN training.
It is demonstrated from the comparisons of the deep CNN method  and the conventional methods with the impact parameter-sensitive observables, that the deep CNN method shows better performance for determining the impact parameters, especially leading to the capability of providing better recognition of the central collision events.
With a proper consideration of the experimental filter effect in both training and testing processes to keep consistency with the actual experiments, the good performance of the deep CNN method holds, and shows significantly better in terms of predicting the impact parameters and recognizing the central collision events, compared to that of the conventional methods, demonstrating the superiority of the present deep CNN method.
The deep CNN method  with the consideration of the filter effect is applied in the deduction of nuclear stopping power. Higher accuracy for the stopping power deduction is achieved benefitting
from the better impact parameter determination using the deep CNN method, compared to using the
the conventional methods. This result reveals the importance to select a reliable impact parameter determination method in the experimental  deduction of the nuclear stopping power as well as other observables.
\end{abstract}

\end{frontmatter}

%\pagewiselinenumbers
%\switchlinenumbers
\section{Introduction}
Investigations on  heavy-ion collisions are motivated by the unique opportunity to gain insights into the crucial information in the terrestrial experiments on hot and
dense nuclear matter, i.e., the behavior of nuclear equation
of state (EOS) at high densities governing the compression,
the internal structure and many other basic properties in supernovae and neutron stars etc., and on nuclear reaction dynamics, i.e., the amount of dissipated energy, the amplitude of collective motion,
and the competition between various dynamical mechanisms, etc~\cite{Danielewicz02,Lattimer01,Bethe90}.
Significant progresses in the investigations on the heavy-ion collisions have been achieved in recent years both in the experimental and theoretical works.
In general, the experimental investigations on the heavy-ion collisions are performed for measuring effective observables, i.e., energy and angular distributions of ejectiles, isotopic yield ratios~\cite{Tsang2001}, neutron-proton emission ratios~\cite{Famiano2006}, resonances~\cite{Li2007,Klimkiewicz2007,Trippa2008}, collective flow~\cite{zak2012} and isoscaling ratios~\cite{Xu2000,Tsang2001_1,Tsang2004,Huang2010}, etc. Correspondingly, the theoretical  investigations focus on making use of microscopic theories in attempt to constrain the key parameters, and pursue the reaction dynamical mechanisms in physics via comparing with the measured observables. One of the main processes in prior to performing the comparisons between the experimental data and the theoretical model predictions  is the impact parameter determination for the measured events. The impact parameter, being known primarily in theoretical simulations but not directly accessible experimentally, has a significant impact on the final state particle production, so that the experimental observables can vary significantly depending on the impact parameter even for the collision system at a given incident energy and with a given combination of projectile and target.

Various methods have been proposed to experimentally determine the impact parameters of the heavy-ion collisions
in an event-by-event basis based on impact parameter-sensitive observables obtained from the collisions, i.e., the charge (mass) of largest fragment~\cite{David1995}, the charged particle multiplicity~\cite{zhu1995}, the directivity ~\cite{David1995},  the total transverse  kinetic energy~\cite{Tsang1991}, the ratio of transverse to longitudinal energy in the center-of-mass frame~\cite{Reisdorf97}, and the quadrupole momentum tensor along beam direction~\cite{Bauer88},
as well as the combinations of such observables~\cite{phair1992}, etc.
However, it was pointed out by Bass \textit{et al.}~\cite{Bass96} that  all these methods have one common drawback,  that is, they  are generally optimized at the larger impact parameter range, and tend to break down for very central collisions.
Indeed, as demonstrated in Ref.~\cite{Ogilvie89}, the charged particle multiplicity, the quadrupole momentum tensor along beam direction, and the total transverse momentum from the collisions of $^{40}$Ca+$^{40}$Ca simulated by the FREESCO are less sensitive to the impact parameter at the smaller impact parameter range compared to the case at the larger range.

Recently, Li \textit{et al.} carefully examined the validity of impact parameter estimation using the multiplicity of charged particles from the $^{112}$Sn+$^{112}$Sn collisions at the incident energies of 35, 50, 70 and 120 MeV/nucleon within the framework of the improved quantum molecular dynamics model~\cite{wang02}. They found that the accuracy of the impact parameter estimation at the entire impact parameter range with the multiplicity of charged particles decreases rapidly as the incident energy decreases from 120  to 35 MeV/nucleon~\cite{zhang18}.
In particular, the observed decrease of the accuracy with the incident energy becomes more significant for the central collisions at incident energies below 70 MeV/nucleon, resulting in even larger ambiguities in the comparisons between the experimental data and the theoretical simulations~\cite{zhang18}.
To date, there is still no effective method to estimate the impact parameters of the heavy-ion collisions, in particular those for the central collisions, at the low-intermediate energy range from the Fermi energy (several ten MeV/nucleon) to around one hundred MeV/nucleon with proper accuracy.
More efforts are therefore required to develop novel methods for the impact parameter estimation for the heavy-ion collisions at this incident energy range, and of great interest, to further apply the novel methods to extract  underlined physics in the future~\cite{Borderie08}.

In the past decades, a prodigious rise of machine
learning techniques which lead to a range of numerous developments in the field
of nuclear and high-energy physics has been seen~\cite{Pang18,Carleo19,Rui2020,Song2021,wang2021}. Among them,  Bass \textit{et al.} introduced the concept of machine learning to determine the impact parameters of the heavy-ion collisions at the relativistic energy range at the first time in 1994~\cite{Bass94}. In that work, they found that within the framework of the quantum molecular dynamics with an explicit inclusion of isospin and pion production via the delta resonance~\cite{Hartnack89}, the predicted impact parameters with an artificial neural network (ANN) show nearly one-to-one consistency to the true impact parameters initially set in the model~\cite{Bass94}, demonstrating the applicability of the machine learning method in the impact parameter determination.
Later, Bass \textit{et al.} improved the machine learning method by using the  two-dimensional transverse and longitudinal momentum distributions of all emitted charged particles as the inputs. This improvement yielded significantly better performance of the impact parameter prediction by a factor of two, revealing that the more sophisticated inputs are chosen for training, the better accuracy will be achieved~\cite{Bass96}.

In 2020, Li \textit{et al.} introduced the deep learning technique, which is a novel branch of the machine learning that learns multiple levels of representations from data and is capable of recognizing and characterizing more complex data sets~\cite{LeCun15}, in the task of impact parameter determination.
They applied the deep learning technique with two commonly used algorithms, the convolutional neural network (CNN) and the light gradient boosting machine (LightGBM), to predict the impact parameters using the two-dimensional transverse momentum versus rapidity distributions of protons from the Au+Au collision events at around 1 GeV/nucleon simulated by the ultrarelativistic quantum molecular dynamics~\cite{Li2020}. The accuracy of the impact parameter prediction is further improved compared to that in the work of Bass \textit{et al.}, indicating a superiority of the novel deep learning technique in the impact parameter prediction at the relativistic energies. In particular, as found in Ref.~\cite{Li2020}, the impact parameters prediction for the central collisions is well accomplished similar to the case for the peripheral ones.
Later in 2021, Li \textit{et al.} continued to compare the  performance of  ANN, CNN, and  LightGBM in terms of predicting the impact parameters~\cite{Li2021}, in attempt to provide reliable impact parameter determination method for the $^{132}$Sn+$^{124}$Sn experiment at  270 MeV/nucleon performed  at the Radioactive Isotope Beam Factory in RIKEN,  Japan~\cite{Jhang2021}. It was found that higher prediction accuracy is achieved when the deep CNN method is used~\cite{Li2021}.
Therefore based on above previous works, it is expected that accurate impact parameter determination for the heavy-ion collisions, especially for the central collisions, at the low-intermediate energies becomes accessible using the deep learning technique as well.

In this article, we focus on pursuing the feasibility of applying the deep learning  with the  CNN~\cite{Li2020} to determine the impact parameters of the heavy-ion collisions at the low-intermediate incident energies from several ten to one hundred MeV/nucleon, using the events simulated by the constrained molecular dynamics (CoMD) model~\cite{Papa01,Papa05}.
Differing from the case at the high energies above several hundred MeV/nucleon, heavy fragments with charge numbers greater than 3 are copiously produced through the complex multifragmentation process in the low-intermediate energy heavy-ion collisions.
CNN is a deep learning architecture inspired by the natural visual perception mechanism of the living creatures~\cite{Gu18}, and is suitable to process tasks such as image and video recognition, and language processing, etc.
To make the deep CNN method to be applicable in the present task, specific considerations  are made in terms of the input selection, the architecture construction and the  training.
The developed deep CNN method is compared to the conventional methods with  the impact parameter-sensitive observables, to establish its better performance for estimating the impact parameters of the heavy-ion collisions at the low-intermediate incident energies in the entire impact parameter range.
With the knowledge that incomplete experimental fragment detection in an event-by-event basis is inevitable due to the limitations of the detection systems, such as the angular coverage and the energy threshold etc., we further investigate the experimental filter influence on the performance of the deep CNN method  in the impact parameter determination. As an actual application, the deep CNN method with a proper consideration of the experimental filter effect is used in the study of nuclear stopping power.

The article is organized as follows. In Sec. 2, an introduction on the event generation is given, followed by the descriptions about the  input selection, the CNN architecture and the CNN training. In Sec. 3, the results are presented and discussed. Summary and perspectives are given in Sec. 4.

\section{Deep learning with convolutional neural network (CNN)}
\section*{2.1 Event generation}
For this work, the CoMD~\cite{Papa01,Papa05} is applied as an event generator.
Two major reasons are considered for the selection of the CoMD. One is that the fermionic nature of the N-body system
as a general condition  ensures that the occupation probability is smaller than 1 in the entire time evolution of the wave packets. The other is that the CoMD is suitable for the present studies involving heavy collision system and requiring large statistics, due to its capability of well reproducing the experimental observables, and its fast performance in practical computation~\cite{Papa01,Papa05}.

Using the CoMD, 300,000 events of the $^{124}$Sn+$^{124}$Sn collisions at the incident energies of 50, 70 and 100 MeV/nucleon are  simulated, respectively. Concerning the fact that the smaller the impact parameter is, the smaller the fraction of collisions is in general simulations, the CoMD events are simulated with a uniform impact parameter distribution in the interval of $b=0-12$ fm, to avoid the potential insufficient training for the central collisions. Here, the maximum impact parameter $b_{max}=12$ fm is taken from the summation of the radii of the projectile and the target nuclei,
$b_{max} = 1.2\times(A_P^{1/3}+A_T^{1/3})$, where $A_P$ and $A_T$ are the masses of the projectile and the target nuclei, respectively.
For the CoMD simulations, a Skyrme interaction with a 200 MeV incompressibility for the effective interaction and the free nucleon-nucleon ($NN$)  cross sections for the $NN$ collisions are used following the previous works in Refs.~\cite{Papa01,Hua12,liu14}. As the fragments are formed in the heavy-ion collisions at the intermediate energies, many of them are in the excited states and undergo sequential decays prior to being measured by the detectors in experiments.
To take into account the sequential decay effect, the time evolution of the wave packets is computed up to 2000 fm/c, permitting the excited primary fragments to de-excite within the framework of the CoMD~\cite{liu14}. The cold fragments at 2000 fm/c are recognized using a coalescence technique with a coalescence radius of 2.4 fm in coordinate space.
Of the simulated 300,000 CoMD events at the given incident energy, 80\% serve as the training samples, and the remaining 20\% serve as the validation samples.
The training and validation data sets are used to optimize the parameters of the CNN (see Sec. 2.3).

\begin{figure*}[tbp]
\centering
\includegraphics[scale=0.45]{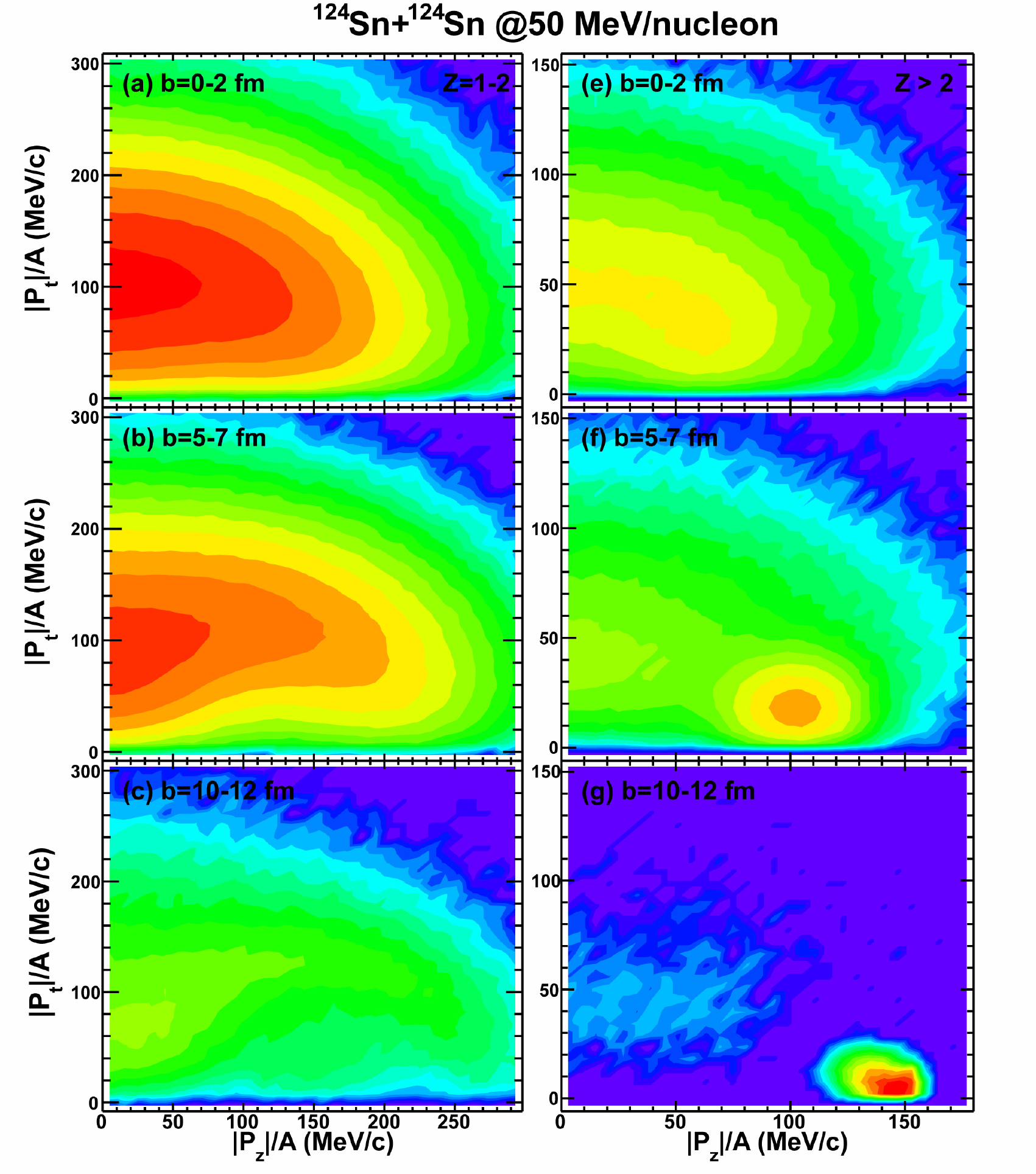}
\includegraphics[scale=0.45]{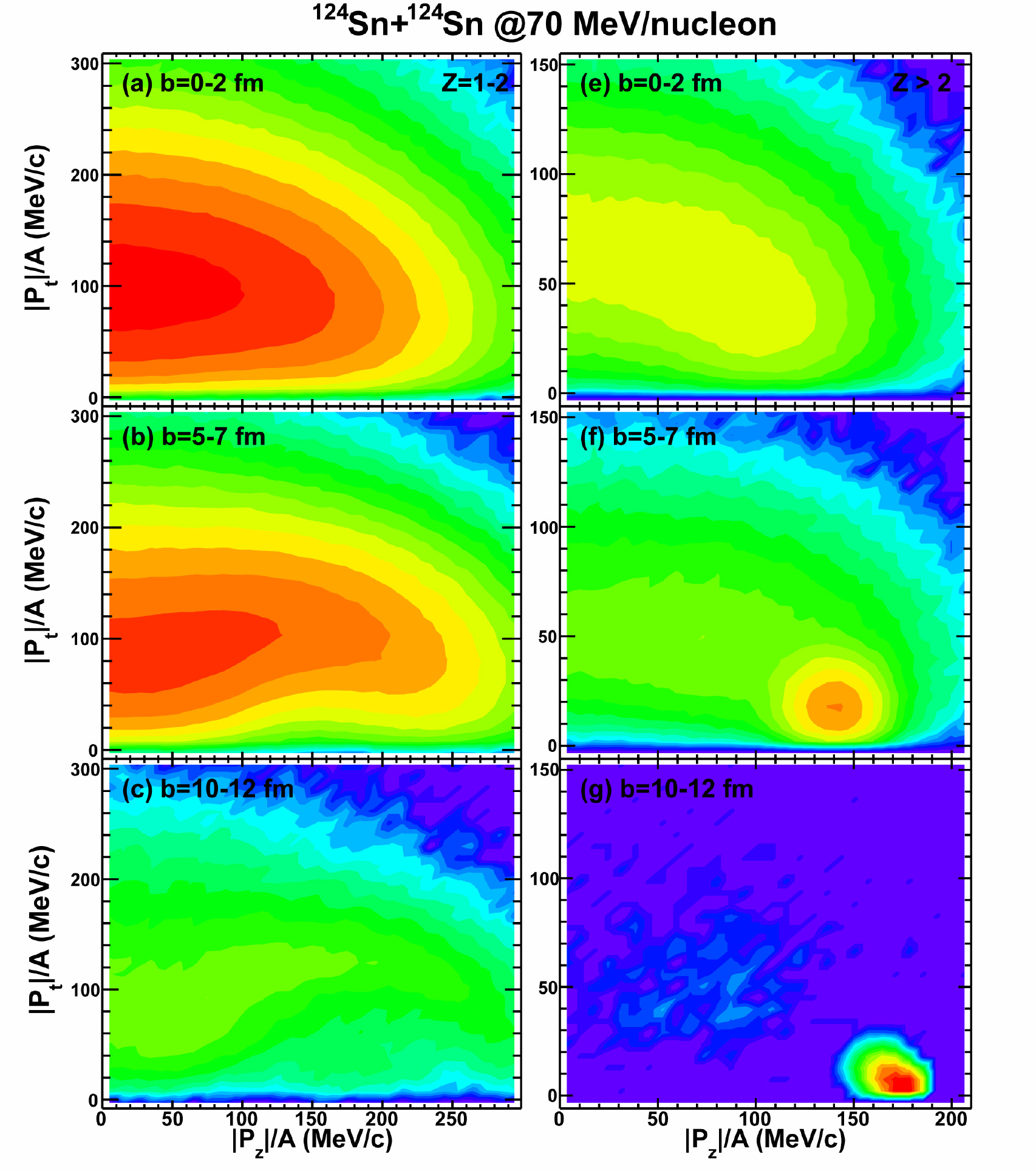}
\includegraphics[scale=0.45]{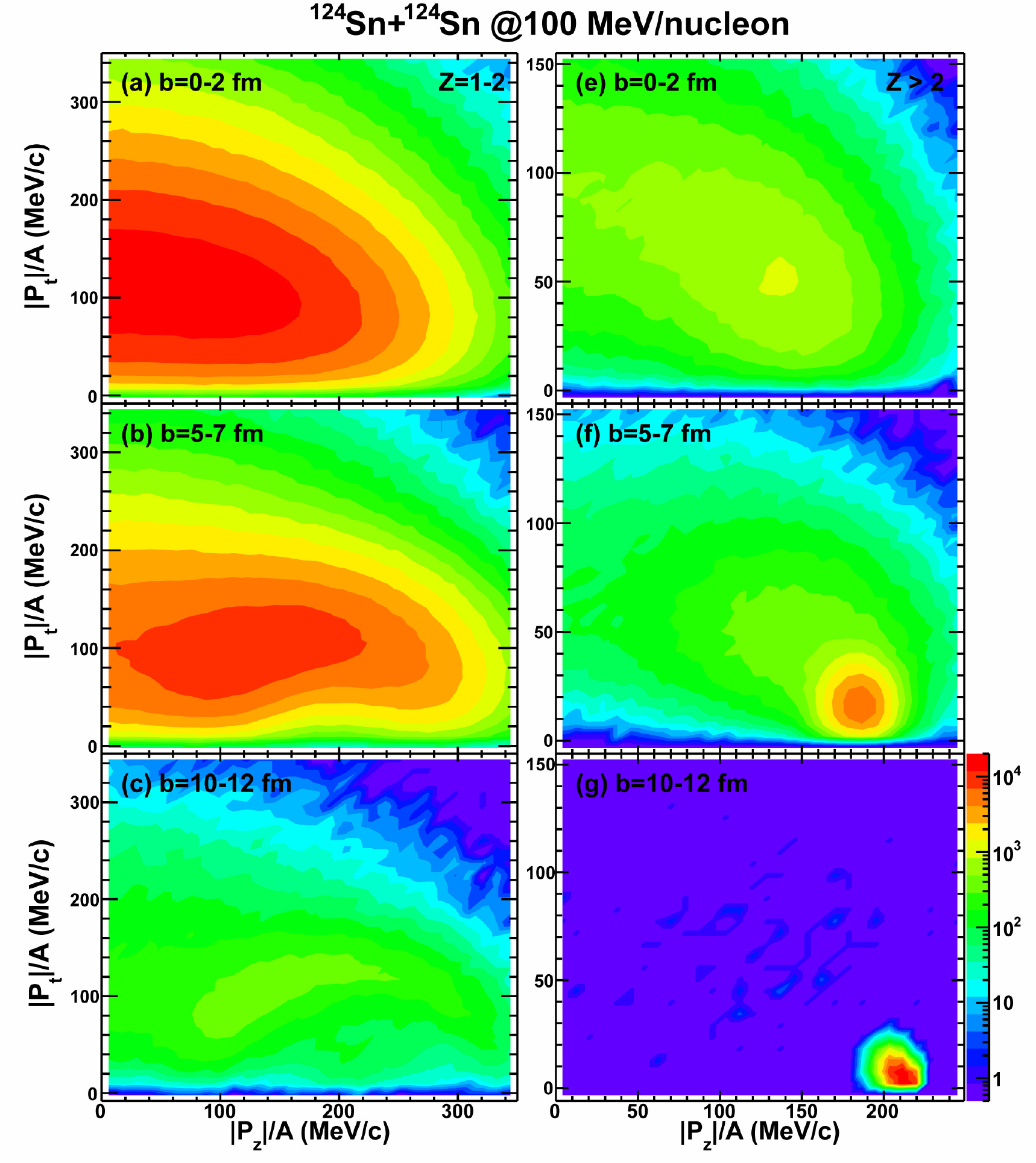}
\caption{\footnotesize  Two-dimensional center-of-mass absolute transverse and longitudinal momentum per nucleon ($|P_z|/A$ versus $|P_t|/A$) distributions of  all available charged particles from the CoMD events of $^{124}$Sn+$^{124}$Sn collisions at 50, 70 and 100 MeV/nucleon.
In each panel,  the results from the impact parameter windows of $0-2$ fm, $5-7$ fm and $10-12$ fm are shown from top to bottom, respectively, and those from light charged particles with $Z\leq2$ and from  heavy charged particles with $Z>2$ are shown on the left and on the right, respectively.
}
\label{fig:fig1}
\end{figure*}

To unbiasedly analyze the trained CNN performance in the actual impact parameter determination, additional 100,000 CoMD events for $^{124}$Sn+$^{124}$Sn at 50, 70 and 100 MeV/nucleon, called the testing samples being independent of the training and validation samples, are, respectively, simulated. For these testing samples, a $bdb$ distribution is taken as such that the probability of an impact parameter $b$ is proportional to $b$ in the interval of $b=0-12$ fm, to keep consistency with the natural case.
Other conditions in the CoMD simulations are set to be identical to those for the training and validation samples.

\begin{figure*}[tbp]
\centering
\includegraphics[scale=0.8]{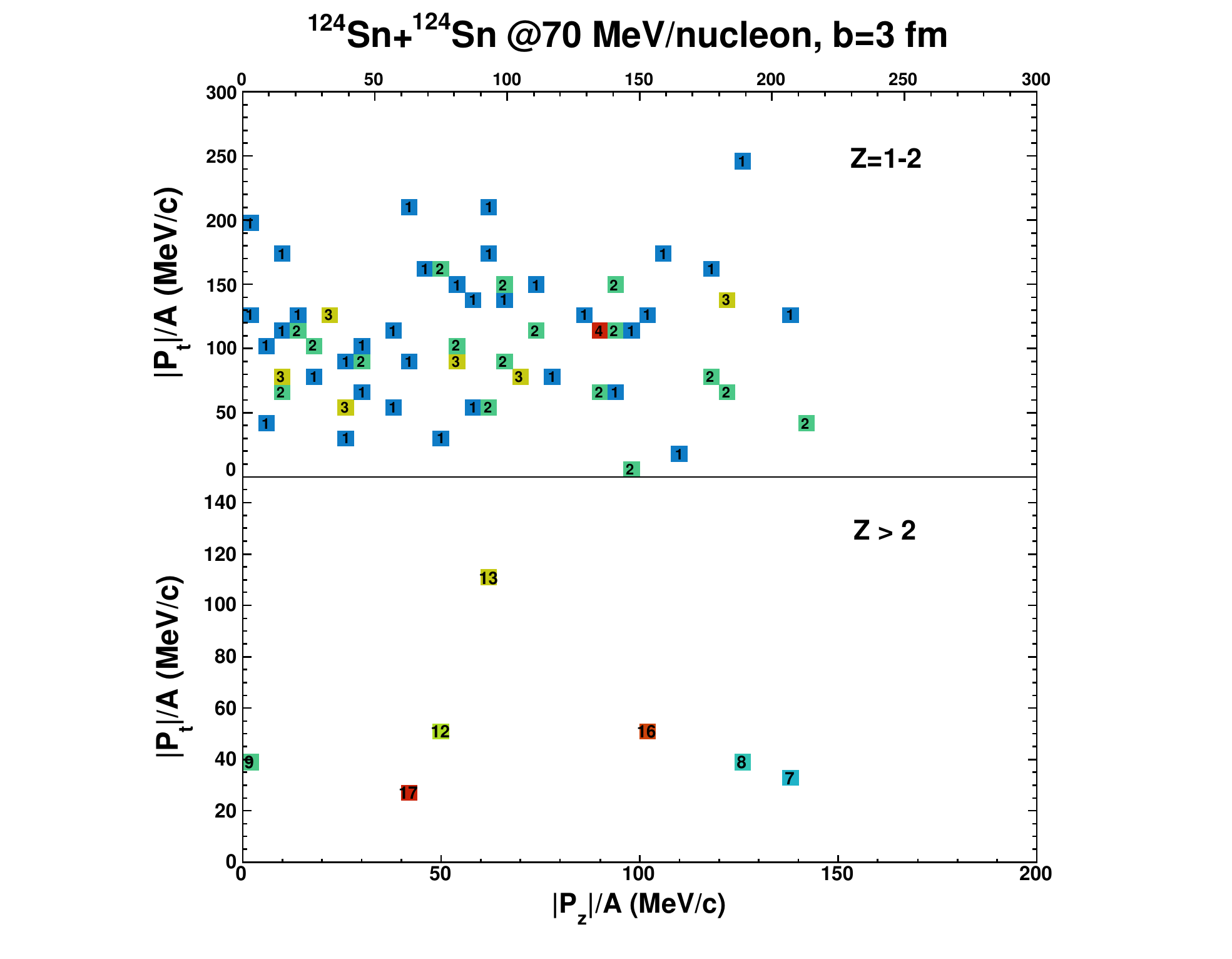}
\caption{\footnotesize  Two-dimensional $|P_z|/A$ versus $|P_t|/A$ histogram  with $50\times50$ pixels for one event  with $b=3$ fm at 70 MeV/nucleon. The upper $50\times25$ pixels with the scales of 0-300 MeV/c along both the $x$-axis and the $y$-axis are  used for light charged particles with $Z\leq2$, and the lower $50\times25$ pixels with the scales of 0-200 MeV/c along the $x$-axis and 0-150 MeV/c along the $y$-axis are used for heavy charged particles with $Z>2$. Each square shown in the figure represents one pixel with  at least fragment entry. The number inside each square represents the summation of nucleons incremented in the pixel.
}
\label{fig:fig2}
\end{figure*}

\section*{2.2 Input selection}

To ensure the accuracy of the impact parameter determination using the deep learning with the CNN, one of the most important steps is to select proper inputs for the training. For this purpose, complex two-dimensional ``images" in the momentum space are adopted.
This input selection  was originally proposed by Bass \textit{et al.}~\cite{Bass96}.
To visualize the  input dependence on the impact parameter learned by the CNN,  the center-of-mass two-dimensional  absolute transverse and longitudinal momentum per nucleon ($|P_z|/A$ versus $|P_t|/A$) distributions for all available charged particles from the CoMD events of $^{124}$Sn+$^{124}$Sn at 50, 70 and 100 MeV/nucleon are plotted in Fig.~\ref{fig:fig1}. In each panel,  the results from the impact parameter windows of $0-2$ fm, $5-7$ fm and $10-12$ fm are shown from top to bottom, respectively, and those from light charged particles with $Z\leq2$ and from  heavy charged particles with $Z>2$ are shown on the left and on the right, respectively.

From the figure, one can observe that the $|P_z|/A$ versus $|P_t|/A$ distributions change significantly with the change of impact parameter selection for all the three incident energies. This can be attributed to the change of the reaction dynamics depending on the impact parameter.
For the central collisions, more $NN$ collisions occur  due to the larger overlap between the projectile and the target, so that more kinetic energies in the longitudinal direction transfer to those in the transverse direction,  resulting in larger $|P_t|/A$ and smaller $|P_z|/A$ for both light and heavy charged particles.
As the impact parameter increases, the energy transfer decreases due to the decrease of the collision violence. Therefore, $|P_t|/A$ becomes smaller, whereas $|P_z|/A$ becomes larger, for both light and heavy charged particles. This significant dependence of the observed $|P_z|/A$ versus $|P_t|/A$ feature on the impact parameter provides  strong support for making use of the $|P_z|/A$ versus $|P_t|/A$ distributions of  the light and heavy charged particles to train the CNN.

\begin{figure*}[t]
\centering
\includegraphics[scale=0.6]{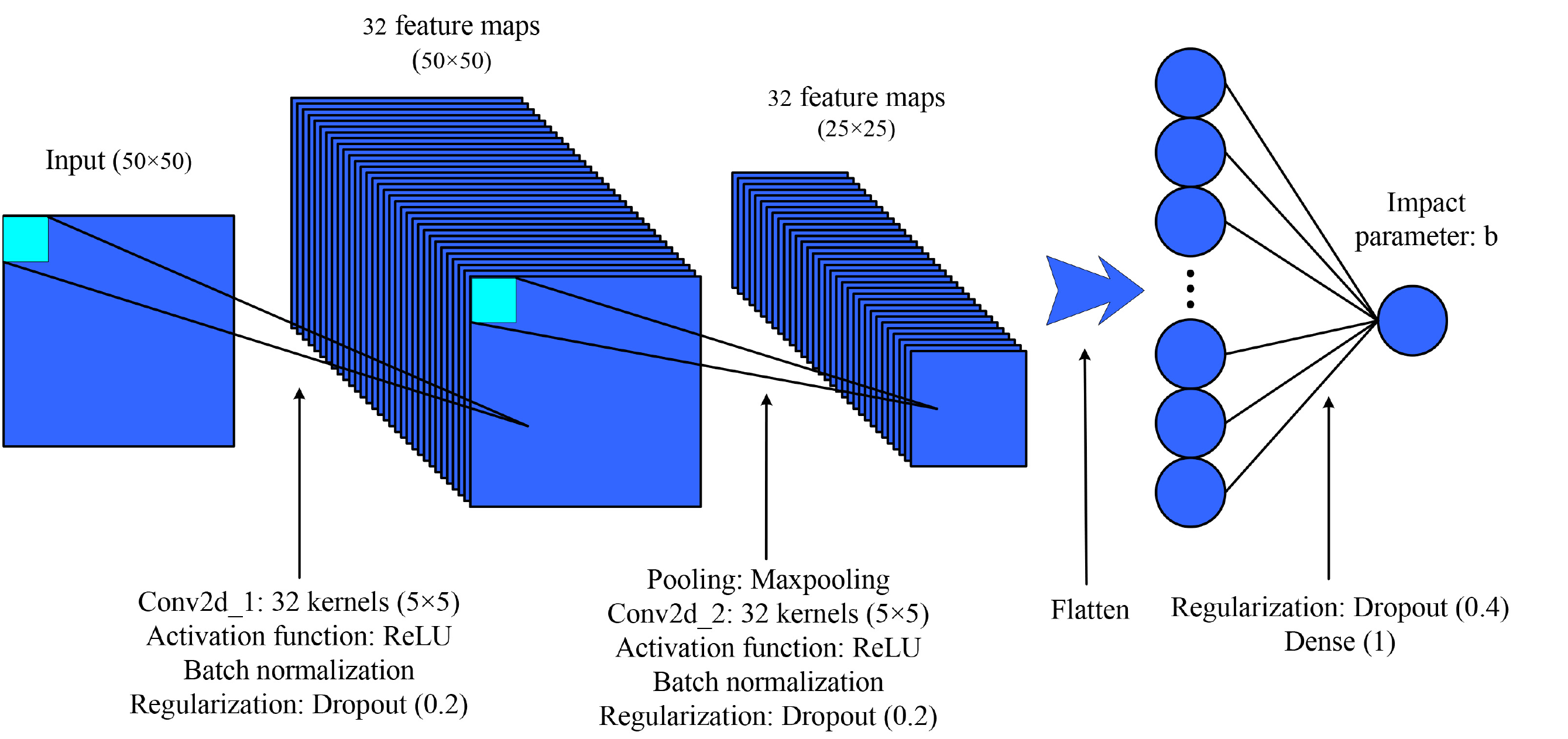}
\caption{\footnotesize  CNN architecture. See the details in the text.
}
\label{fig:fig3}
\end{figure*}

As reflected by the color evolution in the figure, the absolute yields of light charged particles also decrease rapidly, whereas those of heavy charged particles increase rapidly, as the impact parameter increases for all the three incident energies. This may arise a problem of reducing the quality of the CNN training, if one only uses the $|P_z|/A$ versus $|P_t|/A$ distributions from either light or heavy charged particles, due to the small yields for light charged particles at large impact parameters, and for heavy charged particle yields at  small impact parameters  in an event-by-event basis.
To avoid this problem, the momentum information from both light and heavy charged particles are used to generate the inputs for the present CNN training.
It is worthy emphasizing that the problem was unrealized in the previous works of Refs.~\cite{Li2020,Li2021} where only the momentum information of protons were used. However, it is less problematic for those works, since unlike at the low-intermediate energies for the present case,  the yields of light charged particles (such as protons used therein) from one collision event around the relativistic energy range are dominant ejectiles in the entire impact parameter range, and are sufficient enough to train the CNN.

In addition, one may notice that the $|P_t|/A$ and $|P_z|/A$ values for light charged particles both distribute in the  ranges up to $\thickapprox\pm300$ MeV/c, much larger than those of heavy charged particles.
To avoid the possible information missing due to the  overlap of the momentum distributions from the light and heavy charged particles,  an improvement is made by assigning the $|P_z|/A$ and $|P_t|/A$ values of the light and heavy charged particles from one event into two ``images" with different scales, and then combing the two ``images" together as one input for the CNN training.

In Fig.~\ref{fig:fig2},  a typical two-dimensional histogram of $|P_z|/A$ versus $|P_t|/A$ is presented as the input from an event with $b=3$ fm at 70 MeV/nucleon for example. The two-dimensional histogram is set to be with $50\times50$ pixels in total, according to the previous study of Li \textit{et al.}~\cite{Li2020}. In that work, the impact parameter prediction accuracy from the deep learning  with  CNN becomes saturated when the input grids are taken to be larger than $40\times40$.
Using the slightly larger number of pixels is  also considered to account, to some extent, for including the yield information of charged particles. As found below,  the $50\times50$ pixels are large enough to stabilize the impact parameter prediction accuracy for this work.
Of the total $50\times50$ pixels, the upper $50\times25$ pixels with the scales of 0-300 MeV/c along both the $x$-axis and the $y$-axis are  used for light charged particles with $Z\leq2$, and the lower $50\times25$ pixels with the scales of 0-200 MeV/c along the $x$-axis and 0-150 MeV/c along the $y$-axis are used for heavy charged particles with $Z>2$, according to Fig.~\ref{fig:fig1}. Note that the different ranges along the $x$- and $y$-axes for both light and heavy charged particles for the incident energies of 50 and 100 MeV/nucleon are set accordingly.
To further include the particle type information into the input, the histogram is incremented with a weight being the mass number of the given charged particle.
Each square shown in the figure represents one pixel with at least one entry, and the number  inside represents the summation of nucleons incremented in the pixel.  Taking all  pixel contents of the histogram and zero-padding for those with no entry, one can obtain a  $50\times50$ square matrix, namely the CNN input generated from a given event.

\section*{2.2 CNN architecture}
There are numerous variants of CNN architectures in the literature, but their basic components  are similar, including convolutional layer, pooling layer and fully-connected layer in general.
Figure~\ref{fig:fig3} shows the  architecture of the CNN adopted in this work.
As shown in the figure, two convolutional layers are set for learning feature representations of the inputs.
Each convolutional layer is composed of 32 convolutional kernels with size $5\times 5$.
Details about the selection of the number of convolutional layers and the number of kernels are described in the next subsection.
Importing one input to the CNN for example, feature maps are generated from the convolutional layers by first convolving the input with the learned kernels, and then applying nonlinear activation function, batch normalization and regularization on the convolved results.
Here, rectified linear unit (ReLU), known as one of the most notable non-saturated activation functions~\cite{Nair2010}, is taken to introduce nonlinearities into the CNN.
Batch normalization proposed by Ioffe \textit{et al.}~\cite{Ioffe2015} is applied to transform the matrix elements in feature maps to have zero-mean and unit variance.
Regularization is performed using the  algorithm of the dropout function (with rate 0.2) by Hinton \textit{et al.}~\cite{Hinton2012} to effectively reduce the over-training which leads to  model-dependent features and is an unneglectable problem in deep learning with the CNN algorithm. From the first convolutional layer, 32 feature maps are obtained corresponding to the 32 kernels. Then, the 32 feature maps are imported to the second convolutional layer, and are
processed following the similar procedures to generate 32 new feature maps (and so on if there are any more convolutional layers in specific tasks). One pooling layer with the max pooling function are set between the two convolutional layers. The functionality of the pooling layer is to shorten the computational time by reducing the resolution of the feature maps.
As found in the figure, the dimension of the feature maps is reduced by a factor of two from $50\times50$ to $25\times25$ after being processed by the max pooling function.
In the end, the two-dimensional feature maps are flattened to be one-dimensional, and are imported to a fully-connected layer [Dense(1)] following a dropout function with rate 0.4 to transform the extracted features from the convolution into the impact parameter value.

\begin{table*}[tbp]
\caption{Main features of the six types of CNNs  at the incident energies of 50, 70 and 100 MeV/nucleon. The numbers of the trainable model parameters in the  six types of CNNs, which can  reflect their complexity, are tabulated in the third column. The \textit{MSE} values from the validation data sets at the 50th epoch [\textit{MSE}$_{v}(50)$] are tabulated in the fourth column. The mean parallel computational time per epoch on the same Intel i9-10980XE CPU with 18 cores (36 threads) is tabulated in  the fifth column.
} % title of Table
\centering
\begin{tabular}{ p{2.5cm}<{\centering} p{3cm}<{\centering} p{3cm}<{\centering}p{3cm}<{\centering}p{3cm}<{\centering}} %
\hline\hline %
$E_{inc}$ (MeV/nucleon) & Convolutional layer num.$\times$kernel num.  & Model param. num. & \textit{MSE}$_{v}(50)$ (fm$^2$) & Computational time per epoch (sec) \\[.6ex]
\hline
50 & 1 $\times$ 32     & 20933  & 0.76  &  123\\[.6ex]
   & 2 $\times$ 32     & 46615  & 0.60  &  313\\[.6ex]
   & 4 $\times$ 32     &  82561  & 0.55  &  510\\[.6ex]
\cline{2-5}
   & 1 $\times$ 64     & 41756  & 0.78  &  315\\[.6ex]
   & 2 $\times$ 64     &  144279  & 0.59  &  640\\[.6ex]
   & 4 $\times$ 64    &  318497  & 0.55  &  1012\\[.6ex]
\hline
70 & 1 $\times$ 32     & 20933  & 0.55  &  123\\[.6ex]
   & 2 $\times$ 32     &  46615  & 0.43  &  315\\[.6ex]
   & 4 $\times$ 32     &  82561  & 0.40  &  506\\[.6ex]
   \cline{2-5}
   & 1 $\times$ 64     & 41756  & 0.53  &  315\\[.6ex]
   & 2 $\times$ 64     &  144279  & 0.43  &  638\\[.6ex]
   & 4 $\times$ 64    &  318497  & 0.40  &  1025\\[.6ex]
\hline
100 & 1 $\times$ 32     & 20933  & 0.40  &  123\\[.6ex]
   & 2 $\times$ 32     &  46615  & 0.32  &  315\\[.6ex]
   & 4 $\times$ 32     &  82561  & 0.29  &  504\\[.6ex]
   \cline{2-5}
   & 1 $\times$ 64     & 41756  & 0.42  &  315\\[.6ex]
   & 2 $\times$ 64     &  144279  & 0.32   &  640\\[.6ex]
   & 4 $\times$ 64    &  318497  & 0.29   &  1015\\[.6ex]
\hline \hline
\end{tabular}
\label{table:table1}
\end{table*}
\section*{2.3 CNN training}
The training of a CNN is a process of searching for the deep correlation between the input feature representations and the output impact parameter by optimizing the ``model parameters" in the CNN, i.e., the kernel weights and  biases, etc.
The model parameter optimization is always composed of a certain number of epochs, where one epoch is defined as a single training pass through the entire training data set. This training termination at a certain epoch is a general remedy to over-training (if any) in computer science~\cite{Gu18}. See the selection for the number of epochs below.
The measure of how well the CNN has learned the input features during each training epoch is provided by the loss function.
Here, we adopt the mean square error (\textit{MSE}) as the loss function, since the \textit{MSE} loss function performs good and fast, as found by Mallick \textit{et al.}~\cite{Mallick21}. The \textit{MSE} value at a given training epoch (\textit{MSE}$_{t}$) is calculated by
\begin{eqnarray}
MSE_{t}  =  \displaystyle \frac{1}{N_{train}}\sum^{N_{train}}_{i=1} (b_{true,i}-b_{pred.,i})^2,
\label{eq:b_bmax}
\end{eqnarray}
where $b_{true,i}$ is the true impact parameter value for the $i$th training event in the CoMD simulation, and $b_{pred.,i}$ is the corresponding predicted value from the CNN at the given training epoch. The summation runs over the entire training data set with the $N_{train}$ events.
By minimizing the \textit{MSE}$_{t}$ value  epoch-by-epoch, the optimum model parameters can be obtained.
One can expect that the training performance increases as the number of the CNN training events increases, since the CNN is able to gain more  information from more training data. However, the computational time increases significantly as the number of the training events increases simultaneously. We have examined the data size dependence of the CNN training, and found that the training performance becomes saturated when the training event number reaches around 150,000. Therefore for a given incident energy, the 240,000 events used for the CNN training in the present work are enough to ensure the reasonable performance.

Note that in the CNN training, there are also several key hyperparameters being non-trainable but requiring manual tuning, i.e., the number of convolutional layers, the number of kernels, and the epoch number for the CNN training, etc., as mentioned above.
Keeping them in mind, we proceed to discuss the selection of these key hyperparameters, based on the validation data set which is able to provide an unbiased evaluation of a model ``fit" on the training data set while tuning the hyperparameters.

In Refs.~\cite{Li2020,Li2021}, the architecture with 2 convolutional layers and 64 kernels in each convolutional layer (denoted as 2 layers $\times$ 64 kernels, and similarly hereinafter) has been used to establish the CNN to estimate the impact parameters  of the heavy-ion collisions at the high energies from several hundreds MeV/nucleon to 1 GeV/nucleon. Extending their method, six types of CNNs, consisting of 1 layer $\times$ 32 kernels, 2 layers $\times$ 32 kernels, 4 layers $\times$ 32 kernels, 1 layer $\times$ 64 kernels, 2 layers $\times$ 64 kernels and 4 layers $\times$ 64 kernels, are studied as possible candidates. To indicate the complexity of the six CNN architectures, the number of their trainable model parameters are  tabulated in the third  column of Table~\ref{table:table1}.
%Of little surprise is to find in the table that the more complex the CNN architecture is, the more model parameters the CNN has.

The training for the six types of CNNs is performed individually using the training data sets at three incident energies (18 training trials in total), and is commonly computed up to the 50th epoch.  Eighteen sets of CNN model parameters are obtained from the 18 training trials.
The training termination at the 50 epoch for the present CNN training is selected following two considerations. One is that after each training epoch for a given type of CNN at a given incident energy, the obtained CNN is applied to the corresponding validation data set, and the \textit{MSE} value for the validation data set (\textit{MSE}$_{v}$) is calculated following the definition of Eq.~(\ref{eq:b_bmax}). The obtained \textit{MSE}$_{v}$ values are monitored  epoch-by-epoch  together with the \textit{MSE}$_{t}$ values.
Sufficient training for  all the six types of CNNs at  the three energies is evidenced by the fact that for all the training trials, both \textit{MSE}$_{v}$ and \textit{MSE}$_{t}$ decrease epoch-by-epoch, and their decreasing trends  show saturated simultaneously at the 50th epoch. The other is that no cross point is observed between the \textit{MSE}$_{v}$ values and the \textit{MSE}$_{t}$ values as a function of the epoch number, indicating no over-training existing in the present CNN training up to the 50th epoch~\cite{Gu18}.

The obtained \textit{MSE}$_{v}$ values at the 50th epoch
for the six types of CNNs for the three incident energies
are labeled by \textit{MSE}$_{v}(50)$, and are tabulated in the fourth column of Table~\ref{table:table1}.
The \textit{MSE}$_{v}(50)$ values show rather small ranging from 0.29 to 0.78, indicating that the CNN training is well performed for all the 18 trials.
In particular,  the \textit{MSE}$_{v}(50)$ values from the two types of CNNs with  1 convolutional layer show systematically $\approx 30$\% smaller compared to those from the other four types for all the three incident energies.
Concerning the accuracy point of view, one is able to rule out the application of the two types of CNNs with 1 convolutional layer from the present work, due to their larger \textit{MSE}$_{v}(50)$ values.

To select out the suitable one from the remaining four types of CNNs giving similar \textit{MSE}$_{v}(50)$ values, the computational efficiency is  considered as another factor.
In the fifth column of Table~\ref{table:table1}, the parallel computational time per epoch for training the six types of CNNs at the three energies on the same Intel i9-10980XE CPU with 18 cores (36 threads) is  tabulated. One can find that the computational time per epoch increases significantly as the complexity of the CNN architecture increases, where the complexity of the CNN architecture can be reflected by the number of trainable model parameters.
On the other hand, however, the training performance stays more or less same for the remaining four types of CNNs.
This contrast indicates that the saturation of the training performance shows up as the complexity of the CNN architecture reaches to 2 layers $\times$ 32 kernels.
Therefore,  the CNN architecture with 2 layers $\times$ 32 kernels (see Fig.~\ref{fig:fig3}), being the simplest and demanding the shortest computational time in average, is selected for this work. The corresponding three model parameter sets for the incident energies of 50, 70 and 100 MeV/nucleon are saved, respectively, for the following analyses.

\section{Results and discussion}

\section*{3.1 Performance of impact parameter prediction using deep CNN method}
\begin{figure}[t]
\centering
\includegraphics[scale=0.55]{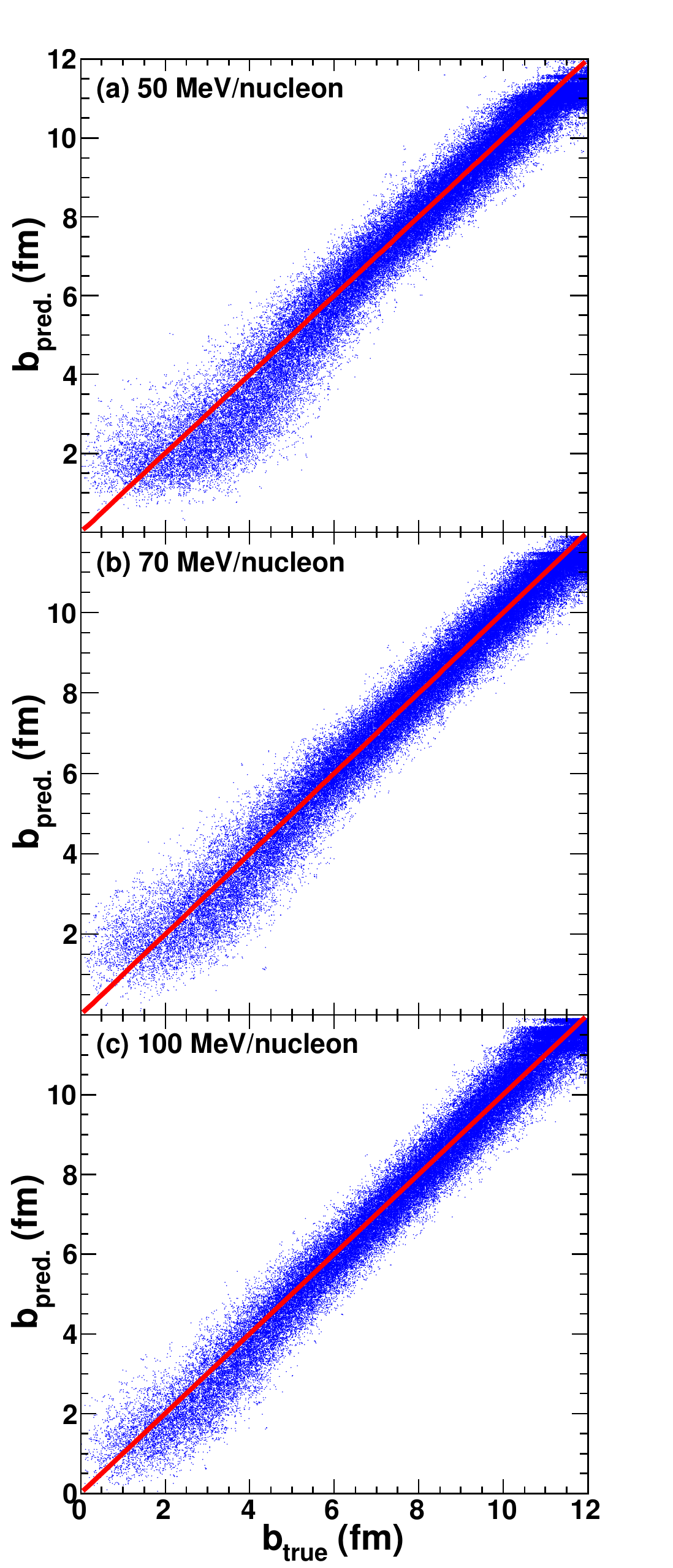}
\caption{\footnotesize  Two-dimensional correlations between  the predicted values of impact parameter using the deep CNN method  ($b_{pred.}$) and their true values ($b_{true}$) from the testing data sets of the $^{124}$Sn+$^{124}$Sn collisions at 50, 70 and 100 MeV/nucleon.  The $y=x$ line in each panel is for guiding the eyes.
}
\label{fig:fig4}
\end{figure}

To  unbiasedly examine the performance of the impact parameter prediction using the CNN from the above training process,
we import the inputs generated from the testing data set at each given incident energy to the trained CNN, and deduce the predicted impact parameter values ($b_{pred.}$) in an event-by-event basis. Figure~\ref{fig:fig4} shows the correlations between the deduced $b_{pred.}$ values from the deep CNN method  and the true impact parameter values ($b_{true}$) set in the CoMD simulations, where the results for the incident energies of 50, 70 and 100 MeV/nucleon are shown in the panels from top to bottom, respectively. The $y=x$ line in each panel is for guiding the eyes.
Most strikingly, as observed in the figure, the data points form narrow bands along the $y=x$ lines in the entire impact parameter range for all the three energies, reflecting a good one-to-one linear relation between  $b_{pred.}$ and  $b_{true}$.

\begin{figure}[tbp]
\centering
\includegraphics[scale=0.4]{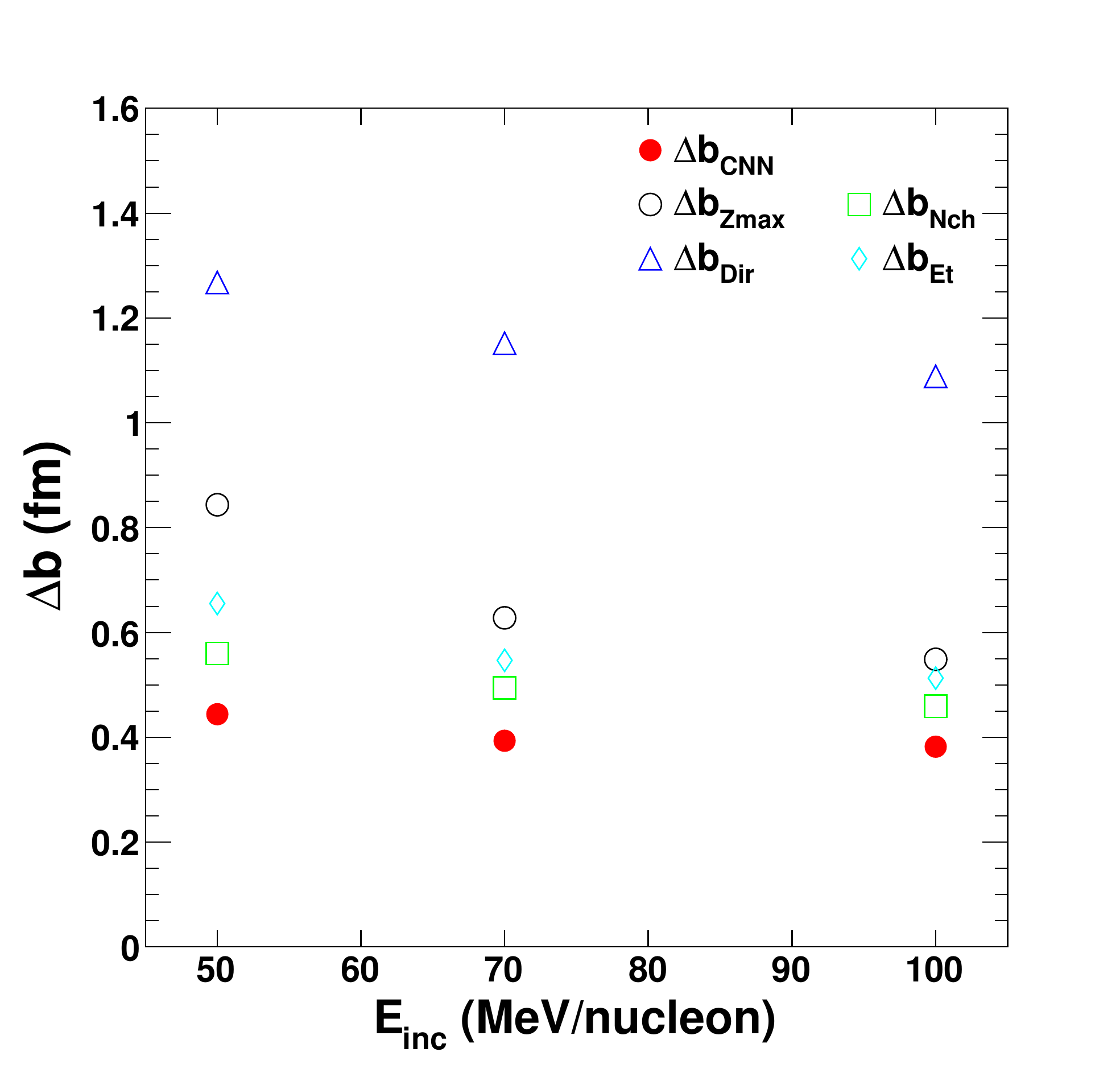}
\caption{\footnotesize
Mean absolute errors defined by Eq.~(\ref{eq:deltab}) as a function of the incident energy. Solid symbols represent the results obtained using the deep CNN method, whereas the open symbols represent those obtained using the conventional methods with the impact parameter-sensitive observables, $Z_{max}$, $N_{ch}$, $Dir$, and $E_t$. See details in the text.
}
\label{fig:fig5}
\end{figure}

To evaluate the accuracy of the impact parameter prediction using the deep CNN method, the mean absolute error ($\Delta b_{CNN}$) is adopted following Refs.~\cite{Bass96,Bass94,Li2020,Li2021}, where $\Delta b_{CNN}$ is deduced by
\begin{eqnarray}
\Delta b_{CNN}  =  \displaystyle \frac{1}{N_{test}}\sum^{N_{test}}_{i=1} |b_{true,i}-b_{pred.,i}|.
\label{eq:deltab}
\end{eqnarray}
The summation in the equation runs over the $N_{test}$ samples in the  testing data set at each given energy. The  $\Delta b_{CNN}$ values for the three incident energies are deduced using the $b_{pred.}$ and $b_{true}$ pairs shown in Fig.~\ref{fig:fig4}, respectively, and are plotted as a function of the incident energy by solid circles in Fig.~\ref{fig:fig5}.
The obtained  $\Delta b_{CNN}$ values, ranging from 0.38 to 0.44 fm, are rather small in magnitude, and these results demonstrate our conjecture, that the impact parameters of the heavy-ion collisions at the incident energies from several ten to one hundred MeV/nucleon can be determined with reasonable accuracy using the deep CNN method, similar to the case at the high energies from several hundreds MeV/nucleon to 1 GeV/nucleon~\cite{Li2020,Li2021}.

\begin{figure}[t]
\centering
\includegraphics[scale=0.55]{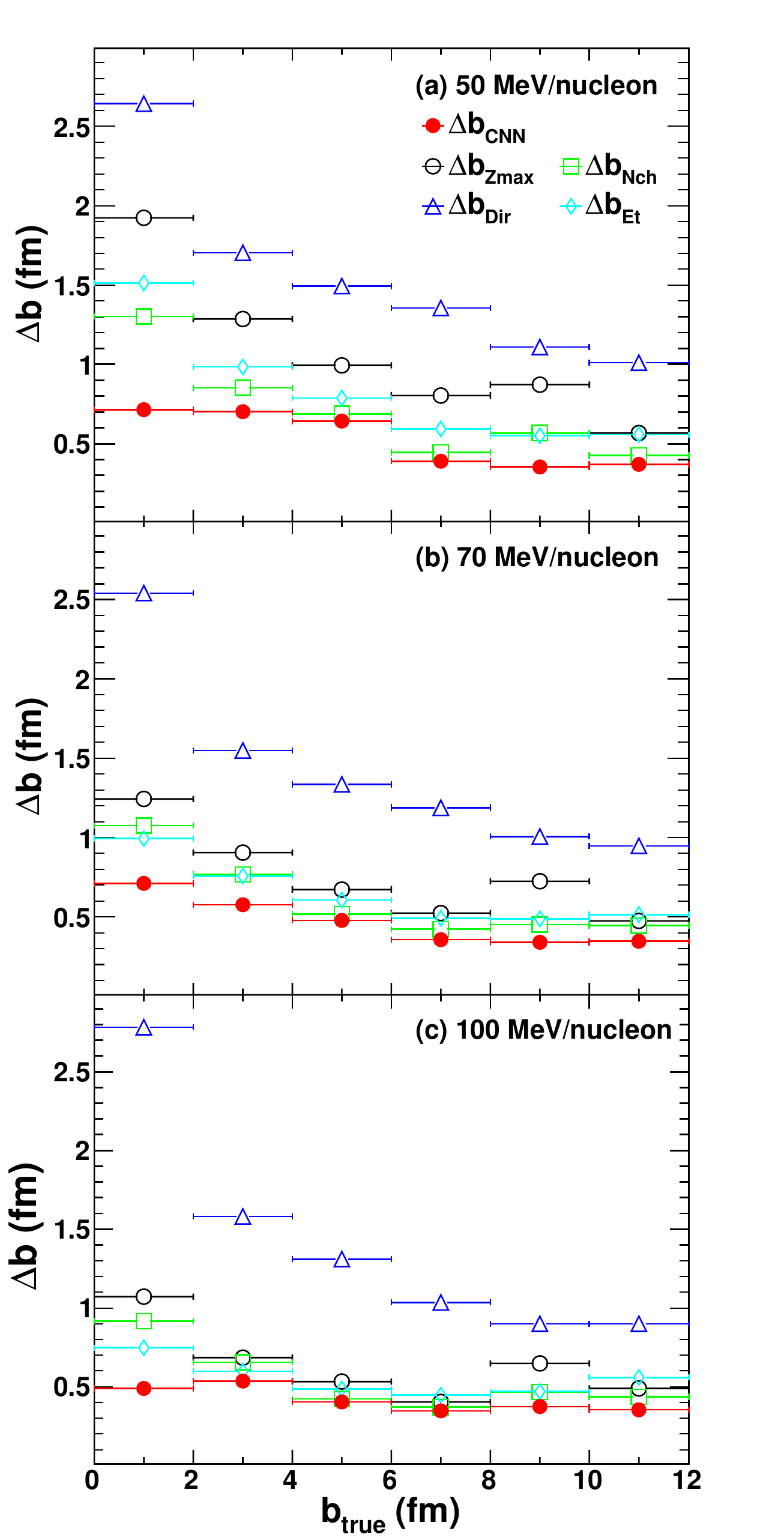}
\caption{\footnotesize
Similar plots as those in Fig.~\ref{fig:fig5}, but as a function of $b_{true}$ and plotted for the incident energies of 50, 70 and 100 MeV/nucleon, separately. The errors shown along the $x$-axis  show the $b_{true}$ intervals in which the $\Delta b_{CNN}$ values are deduced. Symbols are same as those in Fig.~\ref{fig:fig5}. See details in the text.
}
\label{fig:fig6}
\end{figure}

As also observed in Fig.~\ref{fig:fig4},  the linear relation between $b_{pred.}$ and  $b_{true}$ tends to deviate from the $y=x$ line, as $b_{true}$ decreases, commonly for all the three energies.  To gain insights into the dependence of the impact parameter prediction accuracy on $b_{true}$, the $\Delta b_{CNN}$ values are deduced as a function of $b_{true}$ for the three energies, respectively.
The results are shown by solid circles in the panels from top to bottom in Fig.~\ref{fig:fig6}, respectively. In the figure, the decreases of $\Delta b_{CNN}$ as $b_{true}$ increases are more clearly observed for all the three energies, although there is a hint to a slightly weaker dependence for those at the 100 MeV/nucleon incident energy.

The  $\Delta b_{CNN}$ dependence on $b_{true}$ is related to the complex reaction mechanisms  at the low-intermediate energy range.
That is, the low-intermediate energy range is a transition energy range, at which the reaction dynamics is dominated by a mixture effect of the mean field and the $NN$ collisions.
The interplay of the ``attractive" mean field and the ``repulsive" $NN$ collisions imposes large fluctuations in both coordinate and momentum spaces during the time evolution of the nucleus-nucleus collisions.
The large fluctuations  enhance the  characteristic feature similarities among the collisions with different $b_{true}$ values in terms of the emitted particle yield, energy and momentum, etc.,
and further disturb the CNN to learn the characteristic feature representations of the inputs defined using the charged particle center-of-mass transverse and longitudinal momenta in this work.
In particular, as  $b_{true}$  becomes smaller, the fluctuations become larger with the increase of the projectile-target overlap, resulting in more difficulties to distinguish the collisions with smaller $b_{true}$ using the deep CNN method.
Therefore, larger uncertainties are caused in the impact parameter prediction at smaller $b_{true}$ values as found in Fig.~\ref{fig:fig6}.
It should be mentioned that, since the $b_{true}$ values of the testing events are sampled using the $bdb$ distribution, the large uncertainties at smaller $b_{true}$ values lead to small contribution to the ``global" $\Delta b_{CNN}$  presented in Fig.~\ref{fig:fig5}.

Moreover, one may notice that the  $\Delta b_{CNN}$ values in Fig.~\ref{fig:fig5}, as well as those obtained in given $b_{true}$ intervals in Fig.~\ref{fig:fig6}, show a slightly decreasing trend with the increase of the incident energy from 50 to 100 MeV/nucleon. This can be attributed to the simplification of the reaction mechanisms from being dominated by both the mean field and the $NN$ collisions to being dominated by the $NN$ collisions as the incident energy increases. Smaller fluctuations originating from the interplay between the mean field and the $NN$ collisions with the increase of the incident energy lead to better CNN training performance, as indicated by the decrease of  \textit{MSE}$_{v}(50)$ with the incident energy in the fourth column of Table~\ref{table:table1}.
Therefore, higher accuracy of the impact parameter prediction using the deep CNN method  is expected at higher incident energies.

\section*{3.2 Impact parameter prediction for central collisions using deep CNN method}
As highlighted in Ref.~\cite{zhang18} that difficulties exist in determining the impact parameters of central collisions at low-intermediate incident energies using the conventional methods, it is worth studying how well it can be done using the deep CNN method.
In this subsection, we focus on examining  the performance of the impact parameter prediction for the central collisions using the deep CNN method.
In the literature~\cite{Reisdorf97,Lukasik97,Plagnol99,Fevre04}, the central collisions are defined by the interval either with the smallest $b$ or with the smallest reduced impact parameter ($b/b_{max}$), rather than by $b=0$ or $b/b_{max}=0$, since in the strict sense, the collision with $b=0$ fm or $b/b_{max}=0$ has no cross section in nature.  In this work, we adopt the latter, and refer to the collisions with $b/b_{max}<0.2$ as the central collisions following Ref.~\cite{zhang18}.

\begin{figure}[t]
\centering
\includegraphics[scale=0.4]{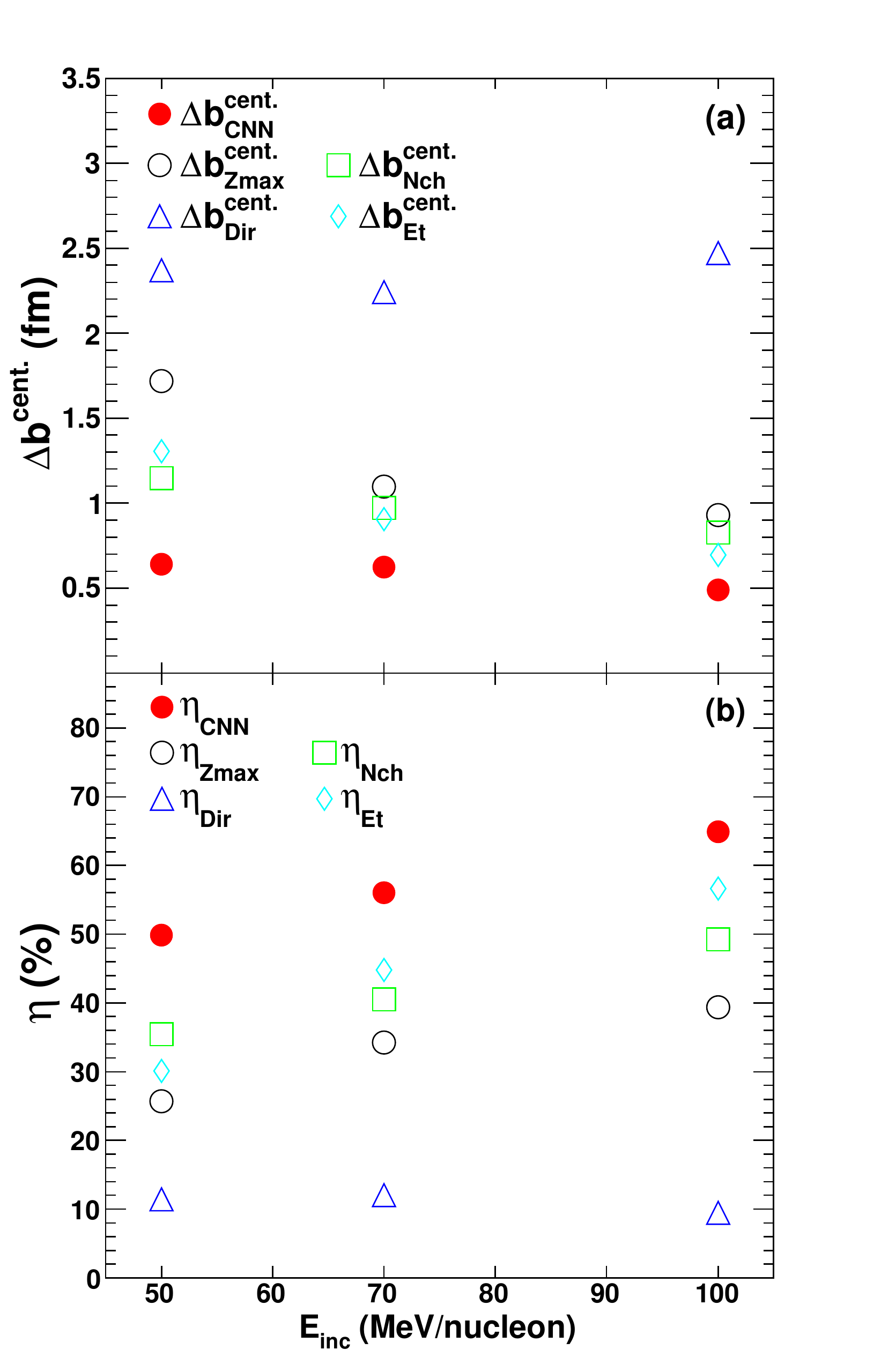}
\caption{\footnotesize
Mean absolute errors  for central collisions  [$\Delta b_{CNN}^{cent.}$, in panel (a)] and  central collision recognition rates [$\eta_{CNN}$, in panel (b)] as a function of the incident energy.  Symbols are same as those in Fig.~\ref{fig:fig5}.
}
\label{fig:fig7}
\end{figure}

Two quantities, the mean absolute error for the central collisions  ($\Delta b_{CNN}^{cent.}$) and the central collision recognition rate ($\eta_{CNN}$), are used to evaluate the impact parameter prediction performance for the central collisions using the deep CNN method. $\Delta b_{CNN}^{cent.}$ is calculated according to Eq.~(\ref{eq:deltab}), but simply limiting the summation up to $b_{true}/b_{max}=0.2$. $\eta_{CNN}$ quantifies the central collision recognition capability, and is defined as
\begin{eqnarray}
\eta_{CNN}  =  \displaystyle \frac{N_{t}}{N_{t}+N_{f+}+N_{f-}},
\label{eq:eta}
\end{eqnarray}
where $N_t$ is the event number of the properly recognized central events with $b_{true}/b_{max}$ and $b_{pred.}/b_{max}$ being both smaller than 0.2. $N_{f+}$ is the number of the false positive events with $b_{true}/b_{max}>0.2$ and $b_{pred.}/b_{max}<0.2$, whereas $N_{f-}$ is the number of the false negative events with $b_{true}/b_{max}<0.2$ and $b_{pred.}/b_{max}>0.2$.

In Figs.~\ref{fig:fig7}(a) and (b), the deduced $\Delta b_{CNN}^{cent.}$ and $\eta_{CNN}$ values from the three testing data sets are presented by solid circles as a function of the incident energy, respectively.
Of little surprise is to find in Fig.~\ref{fig:fig7}(a) that $\Delta b_{CNN}^{cent.}$  shows larger in magnitude but similar in trend as a function of the incident energy, as compared to $\Delta b_{CNN}$ from the entire testing data sets in Fig.~\ref{fig:fig5}.
As also found in Fig.~\ref{fig:fig7}(b), of the $N_{t}+N_{f+}+N_{f-}$ events with $b_{true}/b_{max}<0.2$ or $b_{pred.}/b_{max}<0.2$, more than half can be correctly recognized to be central collision events using the deep CNN method, and the recognition accuracy reaches up to around two thirds at 100 MeV/nucleon.
The increase of $\eta_{CNN}$ as the incident energy increases can be also attributed to
the simpler reaction mechanisms at the higher incident energies.

\section*{3.3 Comparisons between deep CNN method and conventional methods with impact parameter-sensitive observables}
To demonstrate the superiority of the deep CNN method at the present low-intermediate incident energy range, we compare the performance of the impact parameter determination between using
the deep CNN method  and using the conventional methods with impact parameter-sensitive observables in this subsection.
Here, four observables commonly used for experimentally determining the impact parameters of the heavy-ion collisions are taken for comparison, i.e., the charge of the largest fragment ($Z_{max}$)~\cite{David1995},
the charged particle multiplicity ($N_{ch}$)~\cite{zhu1995}, the directivity ($Dir$)~\cite{David1995}, and
the total transverse  kinetic energy ($E_t$)~\cite{Tsang1991}. Having trivial knowledge of $Z_{max}$, $N_{ch}$ and $E_t$, the definition of $Dir$ is given following Ref.~\cite{David1995} as
\begin{eqnarray}
Dir  =  \displaystyle \bigg|\sum_{i}^{N_{ch}} \vec{P_{t,i}}\bigg|/\sum_{i}^{N_{ch}}\bigg| \vec{P_{t,i}}\bigg|,
\label{eq:dir}
\end{eqnarray}
where $\vec{P_{t,i}}$ is the transverse momentum  of the $i$th charged particle, and the summation  is taken over the charged particles in the given event.

\begin{figure*}[tbp]
\centering
\includegraphics[scale=0.55]{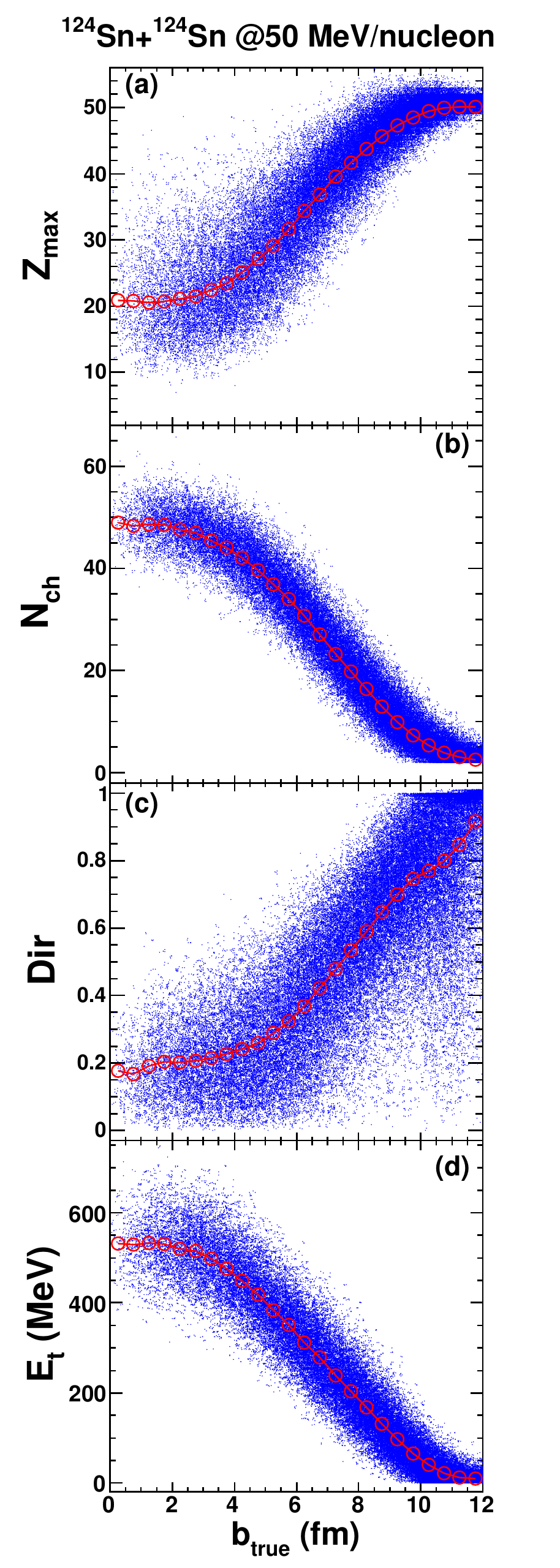}
\includegraphics[scale=0.55]{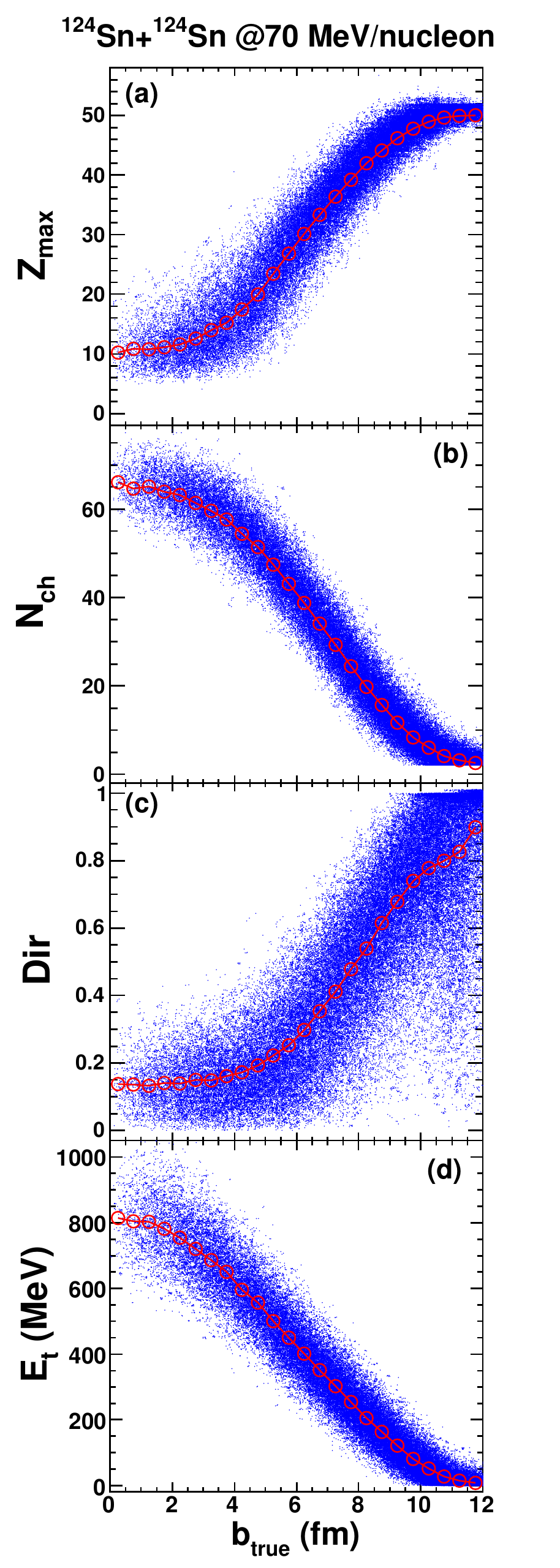}
\includegraphics[scale=0.55]{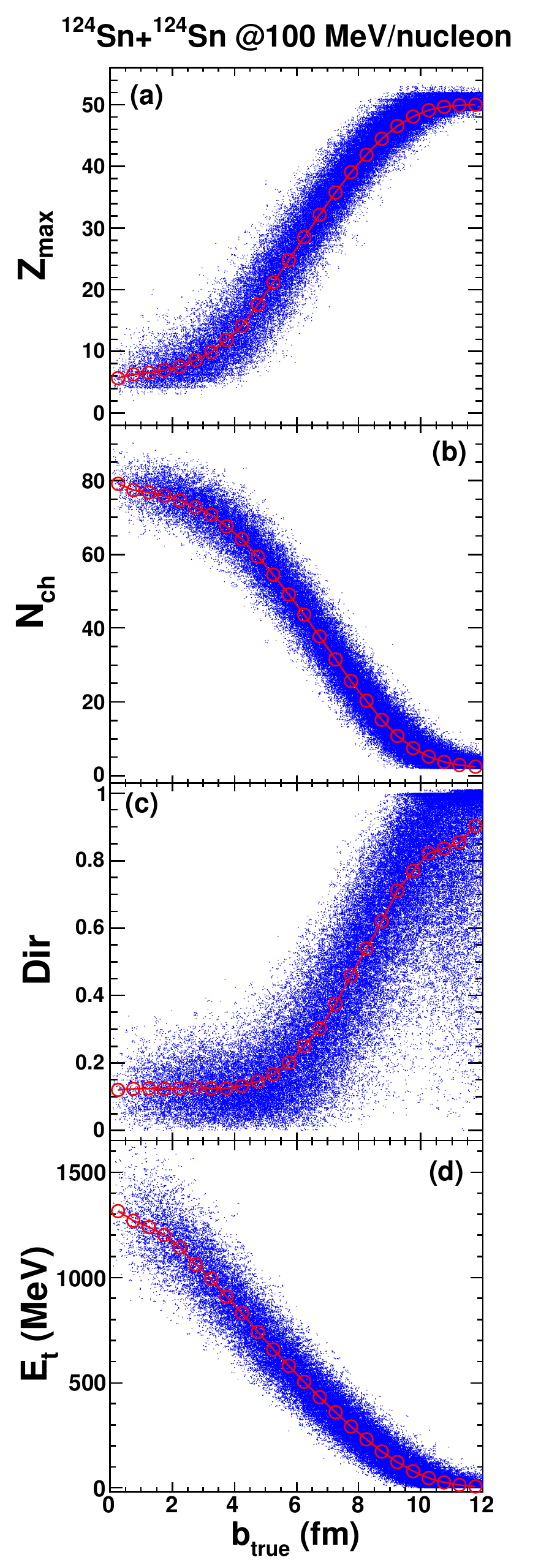}
\caption{\footnotesize
Two-dimensional correlations between the four commonly used the impact parameter-sensitive observables, $Z_{max}$, $N_{ch}$, $Dir$, and $E_t$, and
the $b_{true}$ values from the testing data sets at 50, 70 and 100 MeV/nucleon. The curves with the open circles represent the mean values of the distributions at the given $b_{true}$.
}
\label{fig:fig8}
\end{figure*}

\begin{figure*}[tbp]
\centering
\includegraphics[scale=0.7]{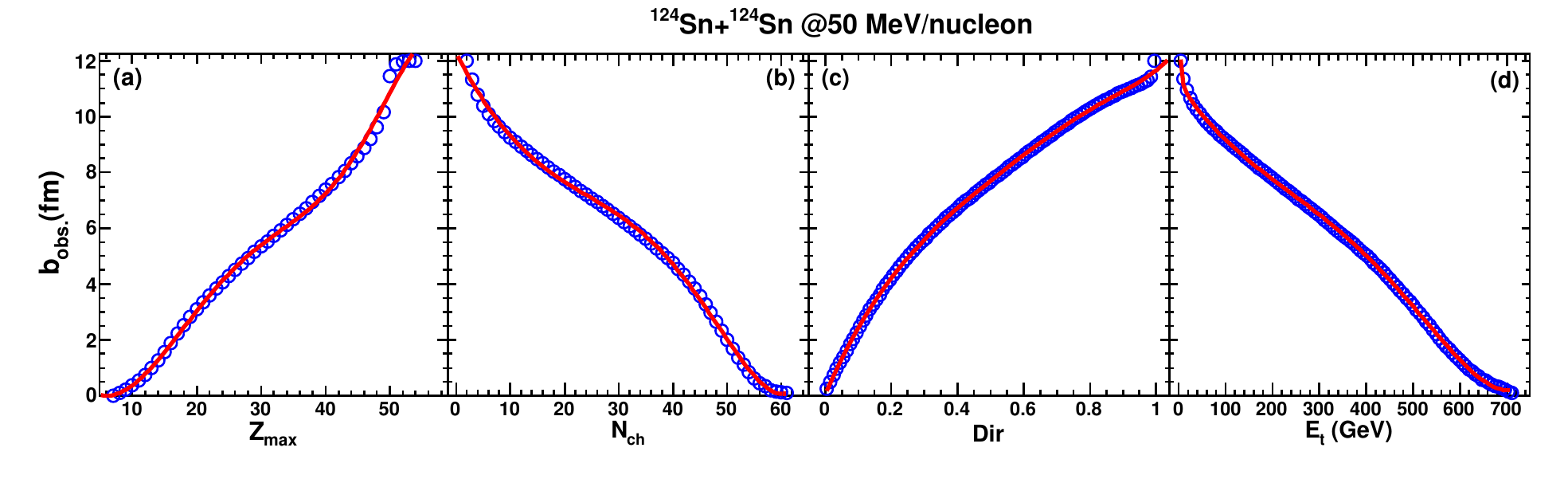}
\includegraphics[scale=0.7]{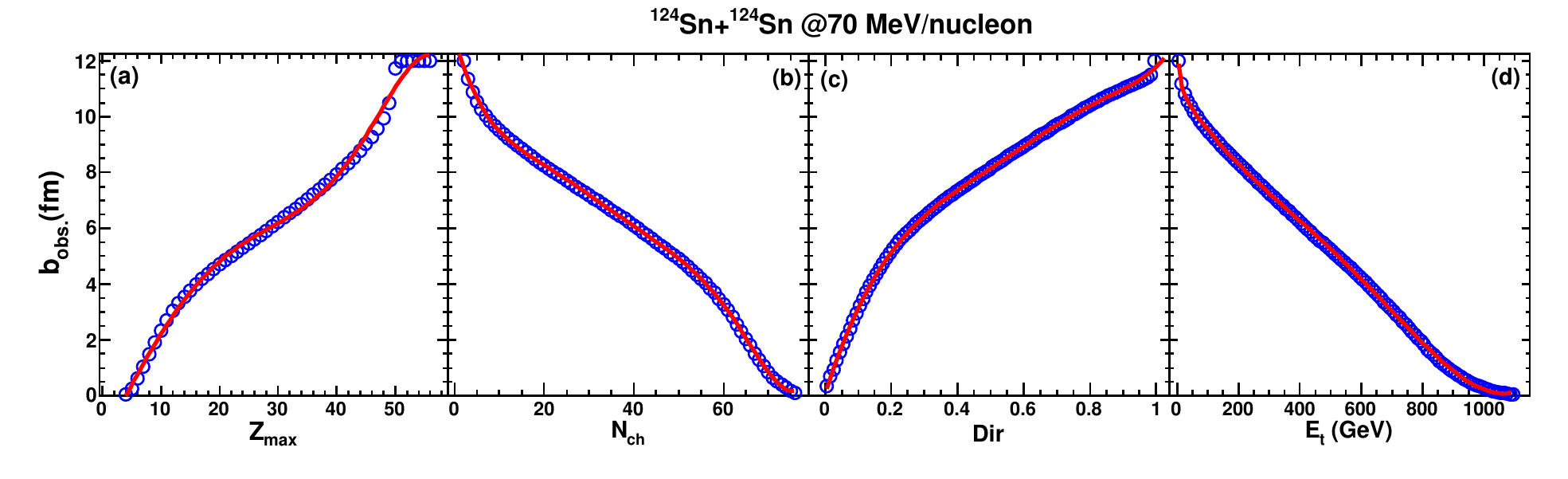}
\includegraphics[scale=0.7]{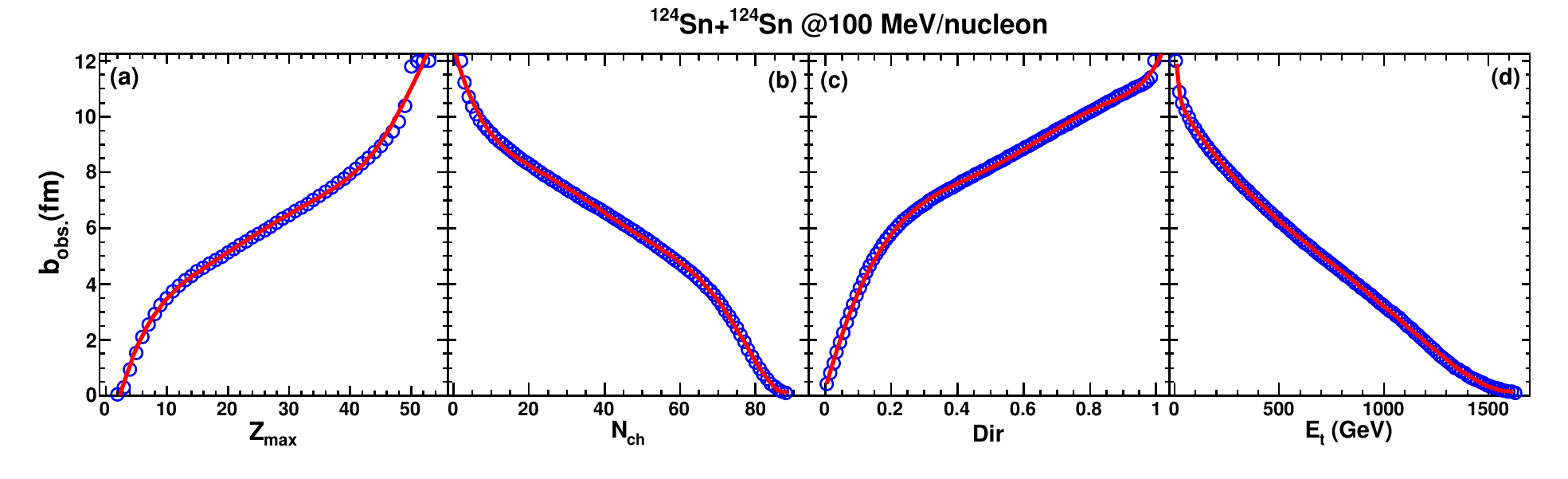}
\caption{\footnotesize
Impact parameter values deduced using Eqs.~(\ref{eq:bobs_Zmax})-(\ref{eq:bobs_Et}) ($b_{obs.}$) as a function of  $Z_{max}$, $N_{ch}$, $Dir$, and $E_t$ for the incident energies of 50, 70 and 100 MeV/nucleon. The curves are the polynomial fits to the data points.
}
\label{fig:fig9}
\end{figure*}

\begin{figure*}[tbp]
\centering
\includegraphics[scale=0.55]{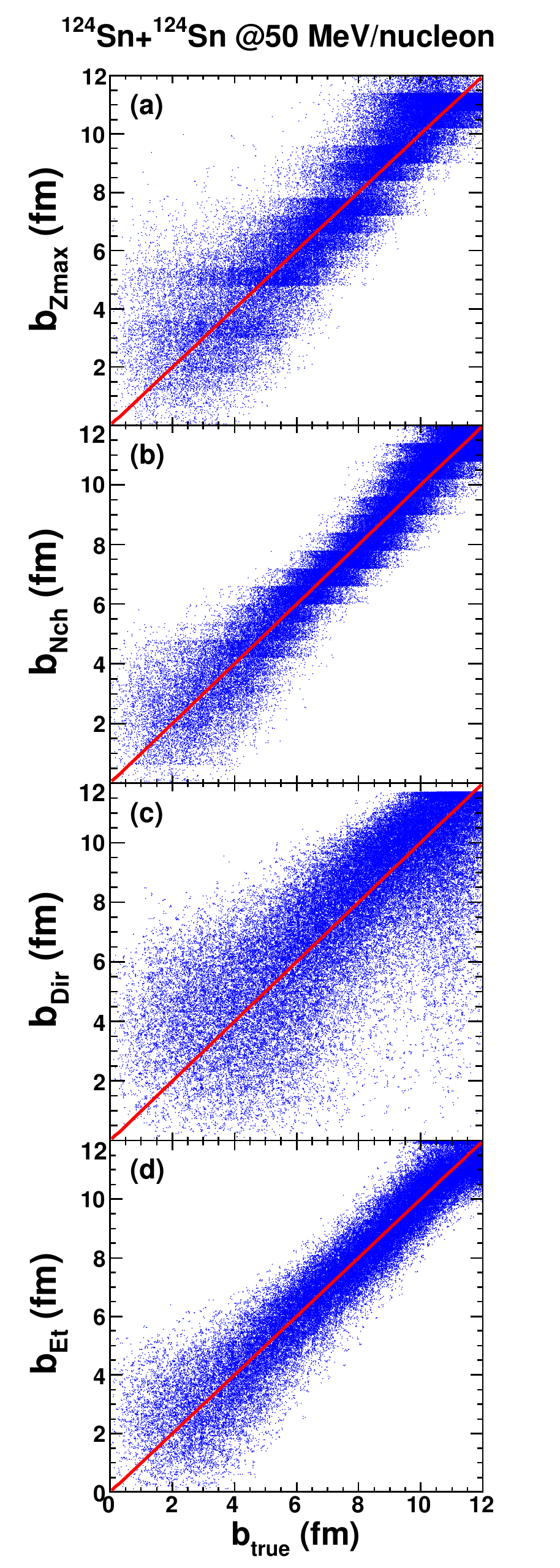}
\includegraphics[scale=0.55]{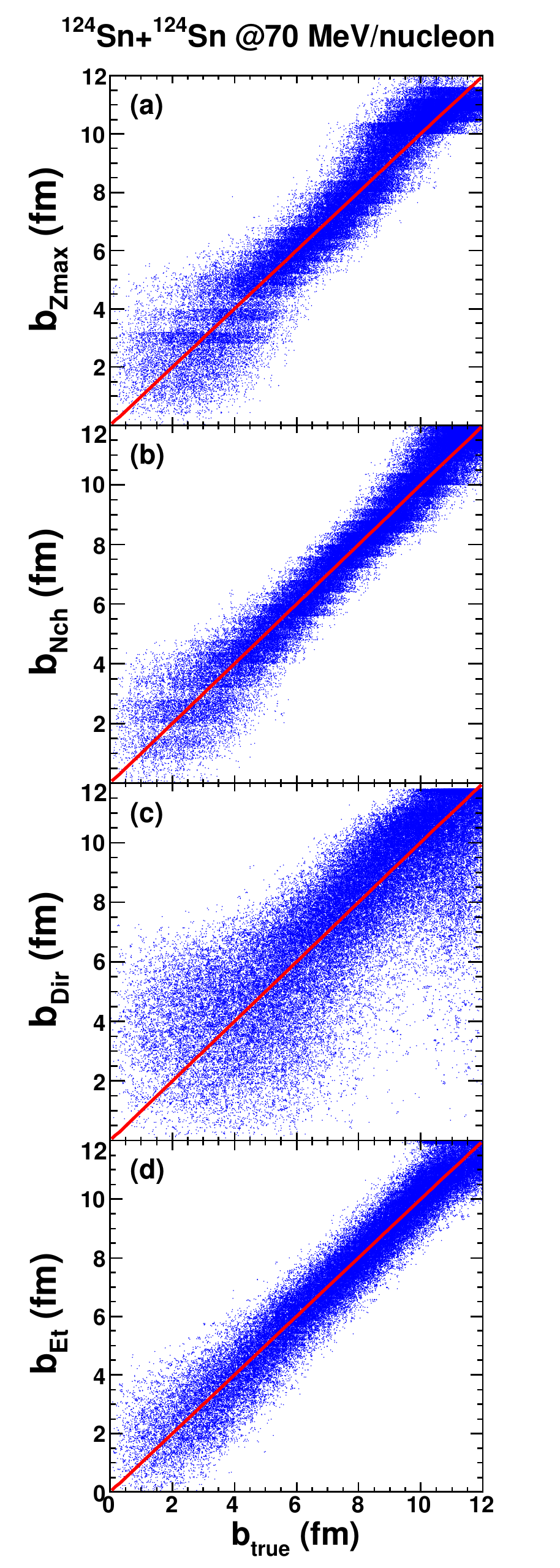}
\includegraphics[scale=0.55]{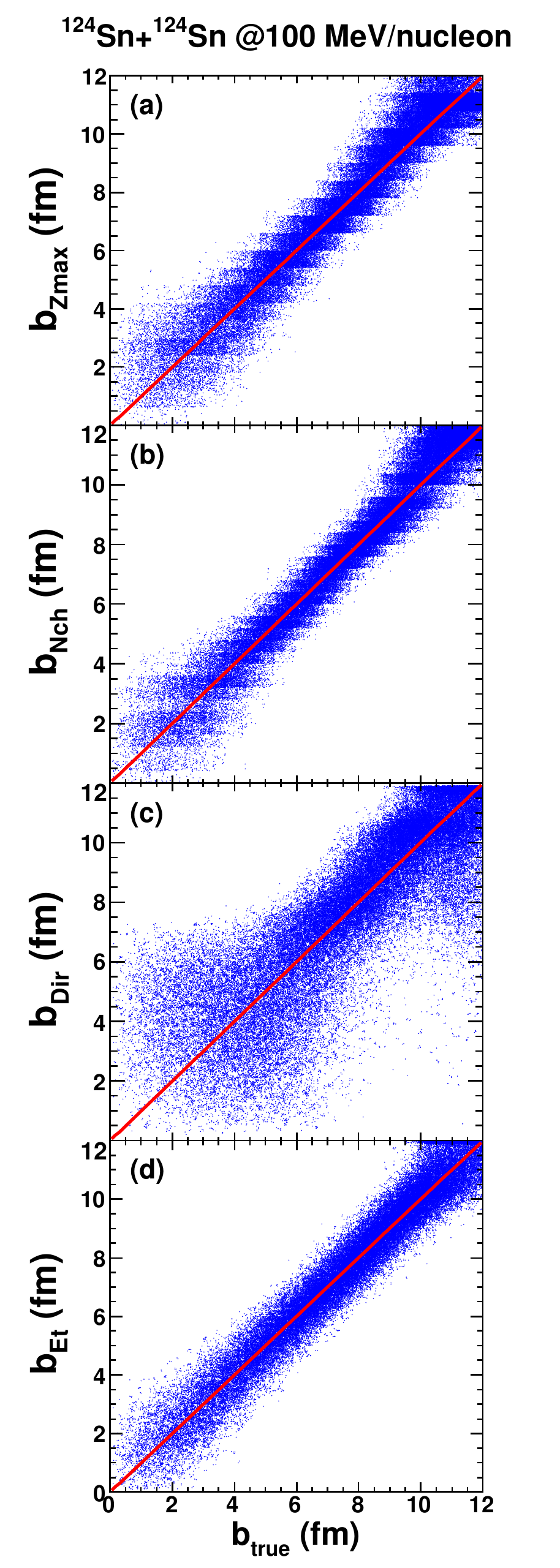}
\caption{\footnotesize
Two-dimensional correlations between the impact parameter values deduced using the four observables (labeled by $b_{Z_{max}}$, $b_{N_{ch}}$, $b_{Dir}$, and $b_{E_t}$ on the $y$-axes) and the $b_{true}$ values from the testing data sets of the $^{124}$Sn+$^{124}$Sn collisions at 50, 70 and 100 MeV/nucleon. The $y=x$ lines are for guiding the eyes.
}
\label{fig:fig10}
\end{figure*}

Figure~\ref{fig:fig8} shows the two-dimensional correlations between the four observables and
$b_{true}$  from the testing data sets
at the incident energies of 50, 70 and 100 MeV/nucleon from left to right, respectively, where the curves with the open circles represent the mean values of the distributions at the given $b_{true}$.  In the figure, in spite of having significantly large fluctuations, all the four observables show a monotonic dependence on $b_{true}$ in the entire impact parameter range.
Making use of the monotonic dependence of the four observables on $b_{true}$, one is able to estimate the impact parameters
directly using the relations below, respectively,
\begin{eqnarray}
b_{obs.}(Z_{max})  =  \displaystyle b_{max}\times \left[\frac{\int^{Z_{max}}_{0}N(Z_{max})\times dZ_{max}}{\int^{\infty}_{0}N(Z_{max})\times dZ_{max}}\right]^\gamma,
\label{eq:bobs_Zmax}
\end{eqnarray}

\begin{eqnarray}
b_{obs.}(N_{ch})  =  \displaystyle b_{max}\times \left[\frac{\int^{\infty}_{N_{ch}}N(N_{ch})\times dN_{ch}}{\int^{\infty}_{0}N(N_{ch})\times dN_{ch}}\right]^\gamma,
\label{eq:bobs_Nch}
\end{eqnarray}

\begin{eqnarray}
b_{obs.}(Dir)  =  \displaystyle b_{max}\times \left[\frac{\int^{Dir}_{0}N(Dir)\times dDir}{\int^{\infty}_{0}N(Dir)\times dDir}\right]^\gamma,
\label{eq:bobs_Dir}
\end{eqnarray}

\begin{eqnarray}
b_{obs.}(E_{t})  =  \displaystyle b_{max}\times \left[\frac{\int^{\infty}_{E_{t}}N(E_{t})\times dE_t}{\int^{\infty}_{0}N(E_{t})\times dE_t}\right]^\gamma.
\label{eq:bobs_Et}
\end{eqnarray}
In Eqs.~(\ref{eq:bobs_Zmax})-(\ref{eq:bobs_Et}), $N(X)$ ($X\in[Z_{max},N_{ch},Dir,E_t]$) represents the event number at the given $X$ value. The exponent $\gamma$ is related to the impact parameter distribution. For the present analysis,  the impact parameters of the collision events in the testing data sets follow the $bdb$ distribution, same as the natural case, so that $\gamma$ is taken to be 1/2~\cite{zhu1995,phair1992,liu14}.

In each panel of Fig.~\ref{fig:fig9}, the impact parameter values deduced from Fig.~\ref{fig:fig8} using Eqs.~(\ref{eq:bobs_Zmax})-(\ref{eq:bobs_Et}) ($b_{obs.}$) are plotted as a function of  $Z_{max}$, $N_{ch}$, $Dir$, and $E_t$ from left to right, respectively.
The relation between the given observable and the corresponding $b_{obs.}$ at a given incident energy can be deduced via the polynomial fits, and the fitting results are shown by curves in the figure.
Using the obtained $b_{obs.}$ versus $Z_{max}$, $N_{ch}$, $Dir$ and $E_t$ relations, we  reconstruct the $b_{obs.}$ values of the testing events in an event-by-event basis, and plot the correlations between the obtained $b_{obs.}$ values and the $b_{true}$ values in Fig.~\ref{fig:fig10}, where the results for 50, 70 and 100 MeV/nucleon are, respectively, shown from left to right with the $y=x$ lines for guiding the eyes.
The impact parameters deduced using the observables $Z_{max}$, $N_{ch}$, $Dir$ , and $E_t$ are labeled by $b_{Z_{max}}$, $b_{N_{ch}}$, $b_{Dir}$, and $b_{E_t}$ on the $y$-axes, respectively.
Note that the jagged structures in the panels (a) and (b) are due to the discreteness of the $Z_{max}$ and $N_{ch}$ observables.
As found in the figure, the data points from the four observables distribute round the $y=x$ lines, but with noticeably wider distributions compared to those of Fig.~\ref{fig:fig4}, for all the three incident energies, indicating poorer accuracy of these conventional methods.

To compare the accuracy of the impact parameter determination using the conventional methods with that of the deep CNN method, we further deduce the quantities, $\Delta b$, $\Delta b^{cent.}$ and $\eta$ following Eqs.~(\ref{eq:deltab}) and (\ref{eq:eta}), using the data points presented in Fig.~\ref{fig:fig10}. The results are plotted in a similar fashion to those deduced using the deep CNN method in Figs.~\ref{fig:fig5}-\ref{fig:fig7}, respectively.
To distinguish from those of the deep CNN method, the results from the four observables are presented by open symbols (see the legends in each figure).
In Figs.~\ref{fig:fig5} and \ref{fig:fig6}, one can find that the $\Delta b_{CNN}$ values as a function of the incident energy and the $\Delta b_{CNN}$ values as a function of $b_{true}$ both show significantly smaller in magnitude compared to those deduced using the conventional methods. This comparison strongly suggests the deep CNN method  has more reliable performance for determining the impact parameters of the heavy-ion collisions at the low-intermediate incident energies than the conventional method with any one of the four impact parameter-sensitive observables.
In particular, as found in Fig.~\ref{fig:fig7}, the $\Delta b_{CNN}^{cent.}$ values are only half to two thirds of the smallest ones obtained from the four conventional observables, and the $\eta_{CNN}^{cent.}$ values show overall significantly larger than the largest ones from the observables at the present energies, clearly demonstrating that the deep CNN method  has capability of providing much higher accuracy in the impact parameter determination for the central collisions at the low-intermediate energy range, compared to the conventional methods.

\section*{3.4 Experimental filter effect on the performance of impact parameter prediction using deep CNN method }
In the following two subsections, we focusing on discussing the application of the deep CNN method for determining the impact parameters from the actual experimental data sets, due to its better performance as evidenced by the quantitative comparisons with that of the conventional methods in the above subsection.
However, in the actual experiments, complete measurement of all the fragments in one event is very difficult, due to the experimental limitations i.e., the angular acceptance of the detector array, and the energy thresholds for the particle detection and identification, etc. It is therefore of great importance to investigate the influence of the experimental filter effect on the performance of the deep CNN method  in prior to applying them to determine the impact parameters of the measured events.

Several 4$\pi$ detector arrays have been built for the investigation on the heavy-ion collisions, i.e.,  the 4$\pi$ array at Michigan State University~\cite{Westfall85},  Miniball/Miniwall array at Laboratoire National SATURNE~\cite{Souza90}, INDRA detector array at GANIL in Caen~\cite{Pouthas95}, NIMROD-ISiS array at Cyclotron Institute, Texas A$\&$M University~\cite{Wuenschel09}, etc.
Since the three incident energies  studied in this work are exactly within the INDRA energy range, we take INDRA as an example to investigate the experimental filter effect on the performance of the deep CNN method  here.

INDRA  is composed of 336 detection modules, arranged in 17 concentric rings covering $2^{\circ}$ to $88^{\circ}$, and $92^{\circ}$ to $176^{\circ}$ in the laboratory frame. $0^{\circ}$ to $2^{\circ}$, and $176^{\circ}$ to $180^{\circ}$ are reserved for the beam pipe, and $88^{\circ}$ to $92^{\circ}$ is shadowed by the target frame. The detection thresholds by the $\Delta E-E$ method for the particle identification are given by the punch-through energies of the front ionization chambers operated at low pressure to be $\approx 1$ MeV/nucleon  for various charged particles.
The charge identification is achieved from hydrogen to uranium, and the isotopic
resolution is achieved up to oxygen, using the $\Delta E-E$ discrimination method~\cite{Neindre02}.  The masses for the heavier fragments with $Z>8$ are estimated from a parametrization of the $\beta$-stability valley. See details about the facility, and basic observables measured using INDRA in Refs.~\cite{Lukasik97,Plagnol99, Pouthas95,Neindre02}.
Here, for keeping consistency with the actual INDRA experiments,  the charged particles in the testing events simulated by the CoMD are filtered using the angular acceptance of the detector array, the detection energy threshold, and the limitation in the charge
identification and the isotopic resolution. Note that the energy threshold filters for the individual detectors are not considered in this analysis, since their effects are negligible, compared to those for the particle identification. In addition, since INDRA only provides charged product information, the neutrons are excluded from the testing data sets.

\begin{figure}[t]
\centering
\includegraphics[scale=0.55]{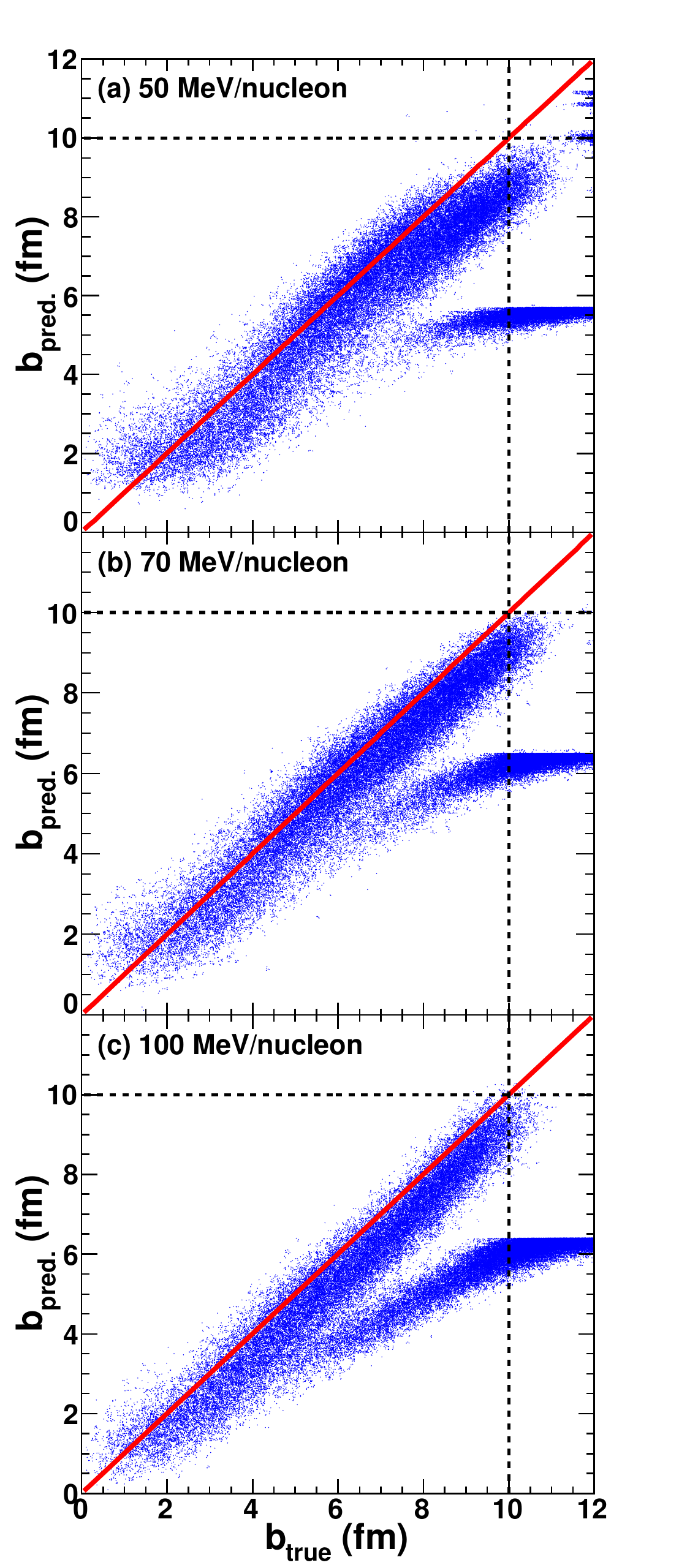}
\caption{\footnotesize   Same plot as Fig.~\ref{fig:fig4}, but from the filtered testing data sets using the deep CNN method  without the consideration of the experimental filter effect.
}
\label{fig:fig11}
\end{figure}

The CNN inputs are regenerated from the three filtered testing data sets, and are imported to the previously trained CNN at the corresponding energies to perform the impact parameter prediction. The correlations between the obtained $b_{pred.}$ values and the $b_{true}$ values are plotted  in Fig.~\ref{fig:fig11}. It is of great surprise to find in the figure that after introducing the filters in the testing data sets, the $b_{pred.}$ values show overall underpredictions with the deep CNN method in the entire $b_{true}$ range for all the three incident energies. Even worse is that two branches of $b_{pred.}$ versus $b_{true}$ data points appear above a certain $b_{true}$ value at the given energy, indicating that the deep CNN method starts breaking down at the $b_{true}$ value. As $b_{true}$ increases to be greater than $\thickapprox 10$ fm, the deep CNN method almost completely losses the impact parameter prediction accuracy, as it hardly provides the $b_{pred.}$ values greater than $\thickapprox 10$ fm.

\begin{figure}[t]
\centering
\includegraphics[scale=0.55]{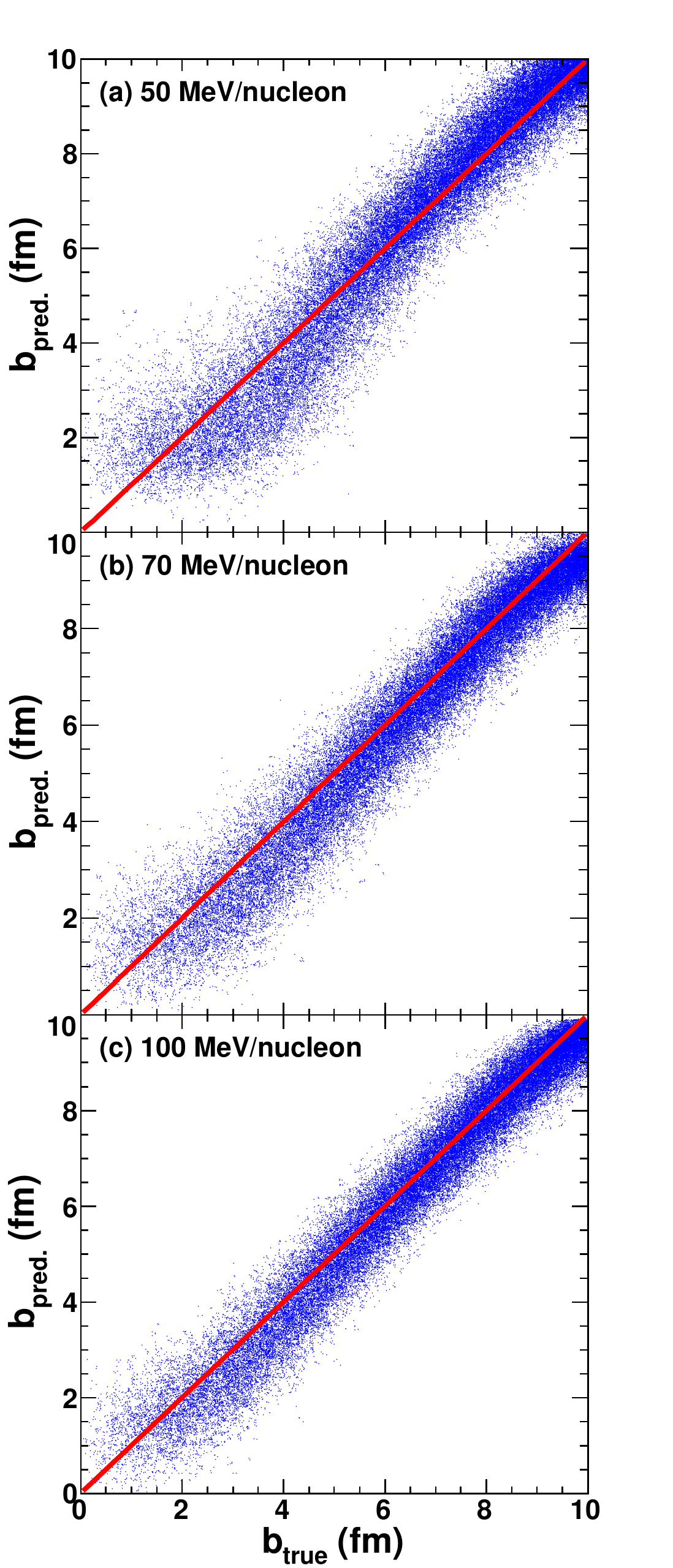}
\caption{\footnotesize
Same plot as Fig.~\ref{fig:fig4}, but obtained from the filtered testing data sets using the deep CNN method with the explicit consideration of the experimental filter effect.
}
\label{fig:fig12}
\end{figure}

These observations can be interpreted by the inconsistency between the input features in the previous training stage and those in the present testing stage, due to the filter limitations in the particle ``measurement" (mostly for the heavy charged particles).
After considering the filter effect in the testing data sets, the collisions with larger $b_{true}$ values, which are supposed to produce abundant heavy charged particles,
appear to produce fewer heavy charged particles, similar to the collisions with smaller $b_{true}$ values. The trained CNN improperly identifies the filtered collision events with larger $b_{true}$ values to be those with smaller $b_{pred.}$ ones in the entire impact parameter range as a consequence.
As $b_{true}$ increases close to the $b_{max}$, the reaction dynamics tends to be binary. One collision event favors to  produce two large charged particles with masses similar to those of projectile and target, namely projectile-like fragment and target-like fragment, as well as a small number of light charged particles.
The projectile-like fragment carries large longitudinal kinetic energy closed to that of the incident projectile in the laboratory frame, and escapes along the beam pipe. The kinetic energy of the target-like fragment is nearly zero in the laboratory frame, and is filtered out by the detection threshold. In such scenario, the key reaction features of the collisions are mostly erased by the filters.
Therefore, the $b_{pred.}$ values for some filtered collision events with the large $b_{true}$ values are even more significantly underpredicted compared to those of the others with the same $b_{true}$ values, leading to the two branches of $b_{pred.}$ versus $b_{true}$ data points in Fig.~\ref{fig:fig11}.
We have crosschecked to the filtered testing data events, and found only quite few light charged particles remained in the filtered events with $b_{true}>10$ fm, commonly for all the three incident energies. This fact can explain why the deep CNN method fails to provide reasonable $b_{pred.}$ values at the extremely large $b_{true}$,  $b_{true}> 10$ fm for the presently studied system of $^{124}$Sn+$^{124}$Sn.

\begin{figure}[t]
\centering
\includegraphics[scale=0.52]{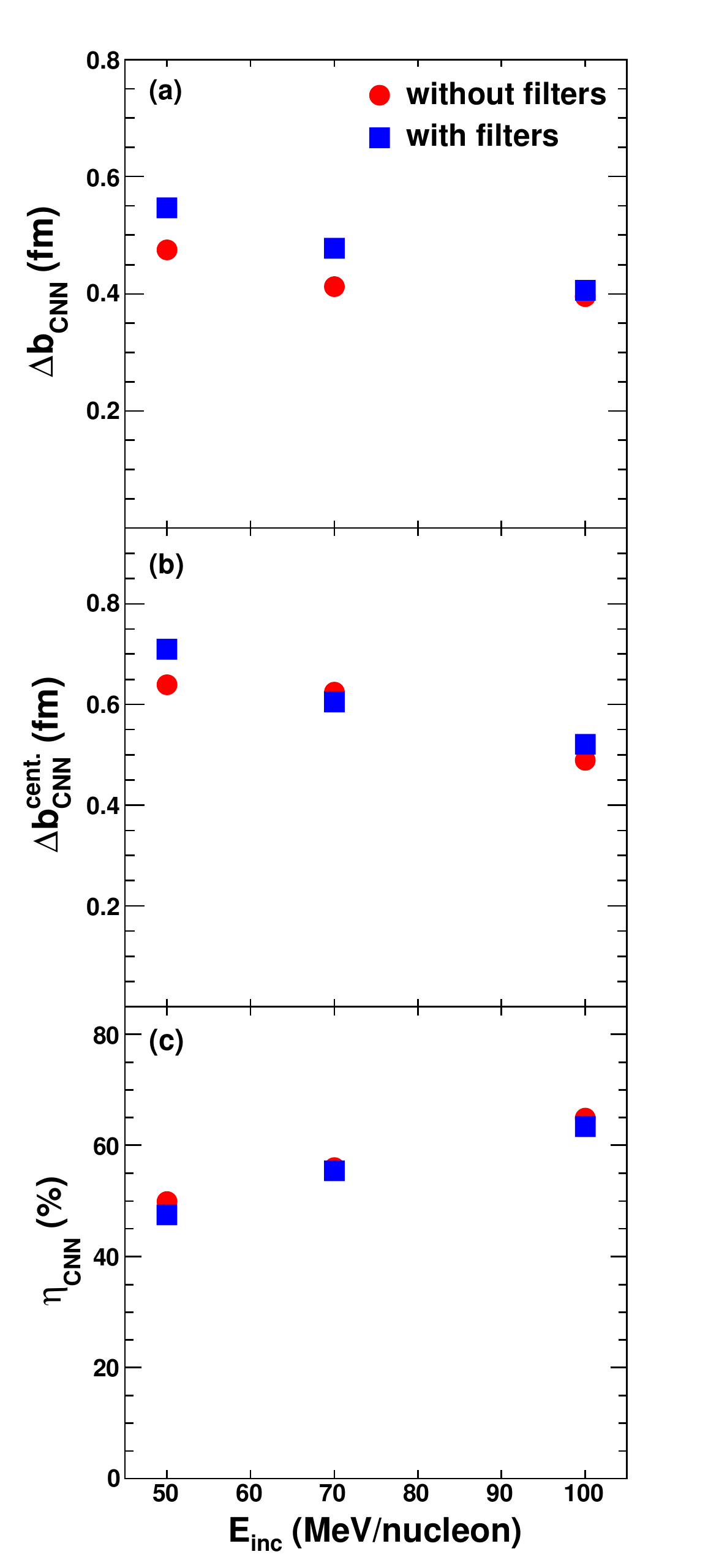}
\caption{\footnotesize
Comparisons of $\Delta b_{CNN}$, and $\Delta b_{CNN}^{cent.}$, and $\eta_{CNN}$ as a function of the incident energy deduced using the deep CNN method  with and without the consideration of the experimental filter effect.
Solid squares and solid circles represent the results deduced from the filtered testing data sets using the deep CNN method with the consideration of the experimental filter effect, and those the unfiltered testing data sets deduced using the deep CNN method without the consideration of the filter effect [having been shown by solid circles in Figs.~\ref{fig:fig5}, and \ref{fig:fig7} (a) and (b)], respectively.
}
\label{fig:fig13}
\end{figure}

\begin{figure}[t]
\centering
\includegraphics[scale=0.52]{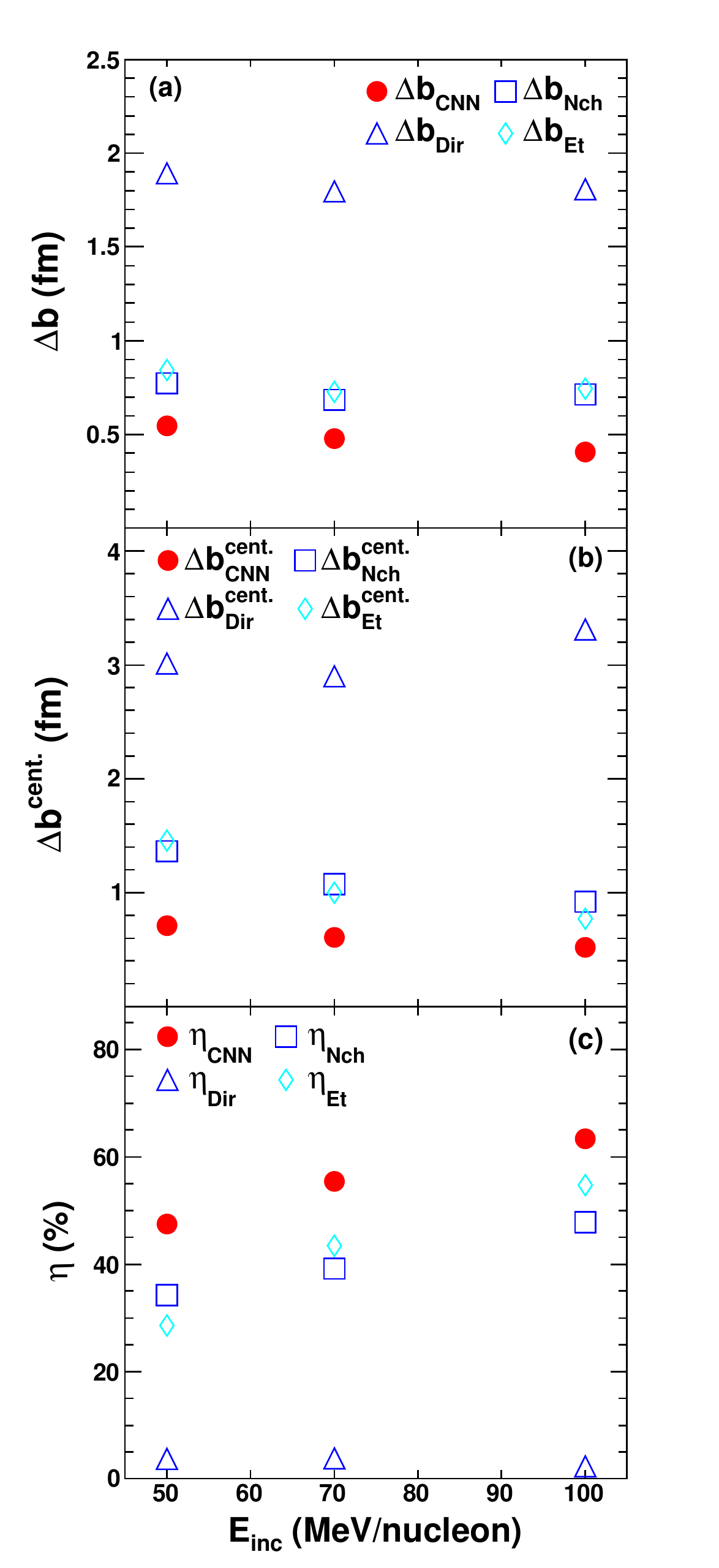}
\caption{\footnotesize
Comparisons of $\Delta b$, $\Delta b^{cent.}$, and $\eta$ as a function of the incident energy deduced using the deep CNN method, and using the conventional methods with the impact parameter-sensitive observables, $N_{ch}$, $Dir$, and $E_t$, with the consideration of the experimental filter effect. Symbols are same as those in Fig.~\ref{fig:fig5}.
}
\label{fig:fig14}
\end{figure}
\section*{3.5 Performance of deep CNN method with consideration of experimental filter effect}
Since the experimental filter effect is far from negligible due to its significant influence on the performance of the deep CNN method, one can not directly apply the presently established deep CNN method for determining the impact parameters of the experimental events, unless one eliminates the input feature mismatch in training and testing the CNN.
For this purpose, an improvement is made by retraining the CNN using the training and validation data sets filtered by the same experimental filters of  INDRA. Additionally as discussed above, the collision events with extremely large $b_{true}$ values at the present studied incident energies are not properly measured by INDRA.  To avoid the potential uncertainties in the CNN training and testing processes caused by this limitation of the INDRA array, we further exclude the events with $b_{true}>10$ fm in both processes. This treatment is reasonable, since these events with $b_{true}>10$ fm are easily filtered out using the total detected charge window in the off-line software~\cite{Lehaut00}.

\begin{figure}[t]
\centering
\includegraphics[scale=0.6]{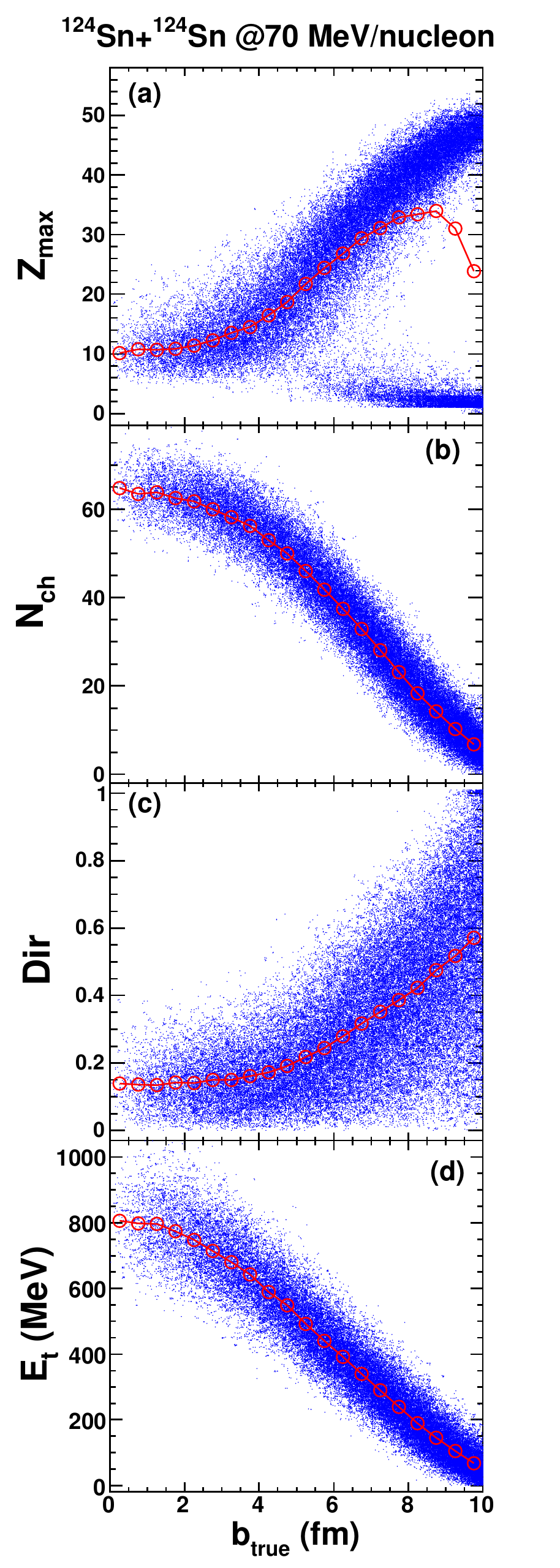}
\caption{\footnotesize
Same plot as Fig.~\ref{fig:fig8}, but only from the filtered testing data sets at the incident energy of 70 MeV/nucleon.
}
\label{fig:fig15}
\end{figure}

Three new model parameter sets for the deep CNN with the experimental filter effect explicitly taken into account are obtained, following the same training procedures presented in Sec. 2 but replacing the original training and validation data sets with the filtered ones at the given energy.  The $b_{pred.}$ values are rededuced using the deep CNN method with the consideration of the filter effect, and the correlations between the obtained $b_{pred.}$ values and the $b_{true}$ values for 50, 70 and 100 MeV/nucleon are plotted with the $y=x$ lines in the panels from top to bottom in Fig.~\ref{fig:fig12}, respectively. One can find that the overall $b_{pred.}$ underpredictions as well as the two-branch features at the large $b_{true}$ disappear in contrast to the results shown in Fig.~\ref{fig:fig11}, indicating a fix of the problem in Fig.~\ref{fig:fig11}.

To further examine whether the accuracy of the deep CNN method in the impact parameter prediction holds after considering the experimental filter effect, we deduce the $\Delta b_{CNN}$, $\Delta b_{CNN}^{cent.}$, and $\eta_{CNN}$ values at the three energies, respectively. The results  are plotted as a function of the incident energy in Fig.~\ref{fig:fig13} (a)-(c), together with those from the unfiltered testing data sets deduced using the deep CNN method without the consideration of the filter effect [shown by solid circles in Figs.~\ref{fig:fig5}, and \ref{fig:fig7} (a) and (b)].
From the comparisons, it is found that the results for both with and without the consideration of  the filter effect are rather consistent with each other both in magnitude and in energy-dependent trend, in spite of being with a tiny deviation in magnitude which indicates a tiny degradation of the accuracy after considering the filter effect leading the partial losses of fragment information.
This fact confirms that the accuracy of the deep CNN method for predicting the impact parameters and recognizing the central collision events holds consistently after considering the filter effect.

For completeness, we deduce the $\Delta b$, $\Delta b^{cent.}$, and $\eta$ values from the filtered testing data sets using the conventional methods following the same procedures in Sec. 3.3, and compare the results in Fig.~\ref{fig:fig14} with those from the deep CNN method with the consideration of the filter effect (shown by solid squares in Fig.~\ref{fig:fig13}).
Note the results from the method with the $Z_{max}$ observable are absent from the figure. This is due to its non-monotonic dependence on $b_{true}$ after considering the filter effect differing from those of the other three as shown
in the top panel of Fig.~\ref{fig:fig15}, where similar plots to those in Fig.~\ref{fig:fig10} are plotted using the filtered testing events, but only for the 70 MeV/nucleon as an example.
As indicated in Fig.~\ref{fig:fig14},  the newly trained deep CNN with the consideration of the filter effect reduces both $\Delta b$ and  $\Delta b^{cent.}$ by $\thickapprox$ 30\% to 50\%, and increases $\eta$ by $\thickapprox$ 20\% to 40\%, compared to the best values achieved by the conventional methods. These comparisons show that, after taking the experimental filter effect in both training and testing processes, the deep CNN method still shows overwhelmingly better performance for predicting the impact parameters compared to the conventional methods. It is a clear demonstration for the superiority of the deep CNN method with a proper consideration of the filter effect in the impact parameter determination for the heavy-ion collisions at the low-intermediate incident energy range, and a strong suggestion for the possibility of its future application in the experimental data analyses.

\section*{3.6 Application of deep CNN method in the study of nuclear stopping power}
\begin{figure*}[t]
\centering
\includegraphics[scale=0.7]{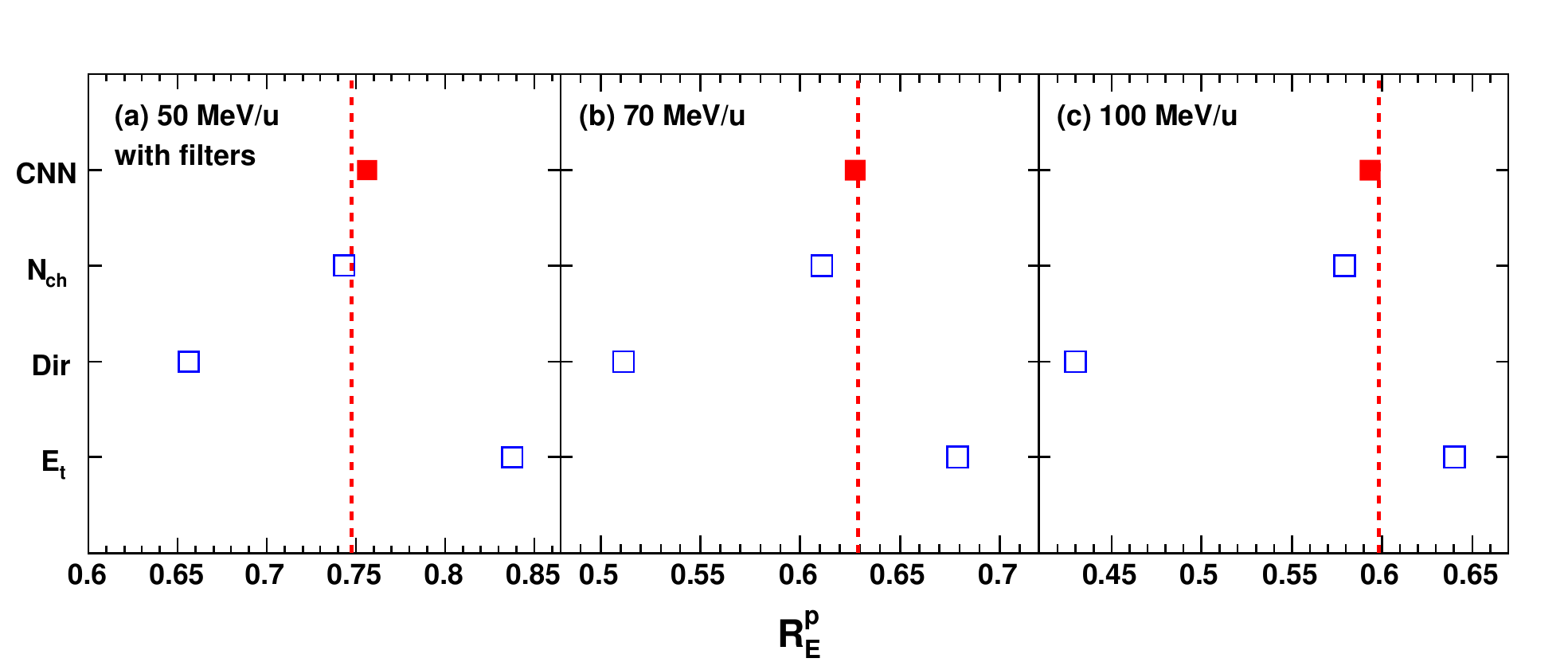}
\caption{\footnotesize
Energy-based isotropy ratios for protons ($R_E^p$) deduced from the filtered ``central" collision events selected using the deep CNN method  with the filter effect (solid squares) as well as using the conventional methods with $N_{ch}$, $Dir$, and $E_t$ (open squares), and their true values from the central events with $b_{true}/b_{max}<0.2$ (dashed lines) as references.
The results in the panels from left to right are for the incident energies of 50, 70, and 100 MeV/nucleon, respectively.
}
\label{fig:fig16}
\end{figure*}

As one of the key observables in the heavy-ion collisions at the intermediate energies, nuclear stopping power provides constraints on the key parameters in nuclear physics, i.e., the nuclear EOS, effective $NN$ interaction and the in-medium $NN$ cross sections, etc., and  helps to elucidate the mechanism of reaction dynamics~\cite{Lehaut00,reisdorf2004,zhang2011,zhao2014,Lopez2014,tian2015,Henri2020}. Since experimental studies of the nuclear stopping power are often performed for the central collisions~\cite{Lehaut00,reisdorf2004,Lopez2014,Henri2020}, the precise impact parameter determination for the central collisions is of great importance. In this subsection, we apply the deep CNN method with the explicit consideration of the experimental filter effect in the deduction of the nuclear stopping power from the filtered testing events.

To quantify the nuclear stopping power, the energy-based isotropy ratio for protons ($R_E^p$) of Henri \textit{et al.}~\cite{Henri2020} is adopted,
\begin{eqnarray}
R_E^p  =  \displaystyle \frac{1}{2} \times\frac{E_{\perp}^{c.m.}}{E_{\parallel}^{c.m.}},
\label{eq:bobs_Rep}
\end{eqnarray}
where $E_{\perp}^{c.m.}$ and $E_{\parallel}^{c.m.}$ are the total center-of-mass transverse and longitudinal energies of the protons from all the selected events.
As indicated in Sec. 2, to accomplish the impact parameter prediction using the deep CNN method, the event-by-event information about the masses, yields, and the momenta for all charged particles is sufficiently proceeded by the complex trained CNN. Therefore, the autocorrelation between the predicted impact parameters, and $E_{\perp}^{c.m.}$ and $E_{\parallel}^{c.m.}$ for constructing $R_E^p$ (if any) is minimized.

The $R_E^p$ values are deduced from the filtered ``central" collision events with reduced impact parameters smaller than 0.2 estimated using the deep CNN method  with the consideration of the filter effect in Fig.~\ref{fig:fig12} as well as from those selected using the conventional methods with $N_{ch}$, $Dir$, and $E_t$, individually. The results are plotted by different symbols in Fig.~\ref{fig:fig16}, from left to right for the incident energies of 50, 70, and 100 MeV/nucleon, respectively. As references, the $R_E^p$ values for the true central events with $b_{true}/b_{max}<0.2$ at the three energies, namely the true $R_E^p$ values, are deduced, and are plotted by dashed lines together in each panel. Note that the statistical errors involved here are rather tiny and are neglected in the figure. As found from the comparison, the $R_E^p$ values deduced from the central collision events selected using the deep CNN method with the filter effect show more consistent in magnitude with the corresponding true $R_E^p$ values for all the three energies.
This comparison indicates that a significant accuracy improvement for the the nuclear stopping power deduction can be achieved using
the deep CNN method with an explicit consideration of the experimental filter effect for the impact parameter estimation, compared to using the conventional methods with the impact parameter-sensitive observables. It also reveals the importance for the selection of the impact parameter determination method in the future experimental investigations on other observables such as fragment energy and angular distributions, isotopic yield ratios, neutron-proton emission ratios, resonances, collective flow and isoscaling ratios, etc.

The present analysis stops here before a further quantitative comparison  between the experimental nuclear stopping power values and the model simulations, as the experimental nuclear stopping power values are deduced based on the impact parameter determination using the conventional method with $N_{ch}$ only~\cite{Henri2020}.
Ever since 1993, a large amount of data have been measured using INDRA  at the low-intermediate incident energy range~\cite{Pouthas95}. We recently started to collaborate with the INDRA group, and the experimental investigation on the nuclear stopping power with the combination of the presently established deep CNN method and the INDRA data is in progress at present.

\section{Summary and perspectives}
In this article, we develop a deep learning based method  with the CNN algorithm for the purpose of determining the impact parameters of the heavy-ion collisions with a reasonable accuracy, especially those of the central collisions, at  the low-intermediate incident energies ranging from several ten to one hundred MeV/nucleon. The collision events of $^{124}$Sn+$^{124}$Sn at 50, 70 and 100 MeV/nucleon simulated using the CoMD are applied, and specific improvements are made in the input selection, the CNN construction and the CNN training. The performance of the established deep CNN method is unbiasedly examined, using the testing data sets independent of the training and validation data sets for training.
The conclusions are summarized as follows:
\begin{itemize}
\item {
The impact parameter prediction accuracy of the deep CNN method in the entire impact parameter range is found to be rather good in the entire impact parameter range, with $\Delta b_{CNN}$ ranging from 0.38 to 0.44 fm for the three incident energies which are comparable to those achieved at the higher energies above several hundred MeV/nucleon in Refs.~\cite{Li2020,Li2021}, demonstrating the feasibility for determining the impact parameters of the heavy-ion collisions at the low-intermediate incident energy range using the  deep CNN method .
}
\item {
The systematic comparisons between the performance of the deep CNN method, and that of the conventional methods with  the impact parameter-sensitive observables, $Z_{max}$, $N_{ch}$, $Dir$, and $E_t$, reveal that the present deep CNN method has capability of providing better accuracy for determining the impact parameters in the entire impact parameter range (for the central collisions in particular), compared to the conventional methods.
}
\item {
The impact of the experimental filters on the performance of the deep CNN method  is carefully investigated using the INDRA detector array as reference, and is found to be rather significant.
After properly considering the experimental filter effect in both training stage and testing stage for keeping consistency with the actual experiments, the good performance of the deep CNN method holds, and shows significantly better in terms of predicting the impact parameters and recognizing the central collision events, compared to that of the conventional methods. These comparisons demonstrate the superiority of the deep CNN method with proper consideration of the filter effect for determining the impact parameters of the heavy-ion collisions at the low-intermediate incident energy range, and leads to a possibility of its application in the further experimental analyses.
}
\item {
The deep CNN method with a proper consideration of the filter effect is applied in the deduction of nuclear stopping power using the CoMD simulated events. Higher accuracy for the stopping power deduction using the deep CNN method  for determining the impact parameters is achieved, compared to those using the conventional methods. The importance to select a reliable impact parameter determination method in the experimental deduction of the nuclear stopping power as well as other observables is suggested.
}
\end{itemize}

As a final remark, the application of the machine learning technique in the impact parameter determination is in the stage of exploration at present. Significant improvements in accuracy and in computational efficiency are expected to be achieved via the algorithm optimization and the novel algorithm application, etc. For example, typical tentative works have been conducted by Sanctis \textit{et al.} ~\cite{Sanctis2008,Sanctis2009}, that a machine learning method based on a novel support vector machine algorithm was first proposed fully independent of those based on the neural networks, and good performance in the impact parameter classification has been achieved as well.
Another crucial issue we should face is that the machine learning requires the aid of theoretical models, so that the model dependent effect is inevitable in the deep learning process. Recently, the transport model comparison project has been proceeded by the researchers around the world in attempt to pursue the ``ideal" transport model, and great progress has been achieved with their concerted efforts~\cite{Xu2016,Zhang2018,Colonna2021}.
It is expectable that the model dependent effect can be almost eliminated adopting the ``ideal" transport model in the deep learning in the future.

\section{Acknowledgments}
This work is partly supported by the National Natural Science Foundation of China (No.
11705242, No. 11805138, No. 11905120, No. 11975091, No. 11947416), the Fundamental Research
Funds for the Central Universities (No. YJ201954 and No. YJ201820) in China, and International Visiting Program for Excellent Young Scholars of Sichuan University. This work is also supported by the US Department of
Energy under Grant No. DE-FG02-93ER40773

\end{document}